\documentclass[useAMS,usenatbib]{mnras} 
\usepackage{graphicx}
\usepackage{color}
\usepackage[usenames,dvipsnames]{xcolor}
\usepackage{mathrsfs}
\usepackage{lscape}
\usepackage{txfonts}
\usepackage{epstopdf}
\usepackage{pifont}

\definecolor{lightgreen}{RGB}{173,221,142}
\definecolor{mediumgreen}{RGB}{65,171,93}
\definecolor{darkgreen}{RGB}{35,132,67}
\definecolor{mintgreen}{RGB}{0,234,128}
\definecolor{lightblue}{RGB}{158,202,225}
\definecolor{mediumblue}{RGB}{66,146,198}
\definecolor{darkblue}{RGB}{33,113,181}
\definecolor{lightorange}{RGB}{254,178,76}
\definecolor{darkorange}{RGB}{255,127,0}
\definecolor{mediumred}{RGB}{238,64,0}
\definecolor{jazz}{RGB}{165,11,94}
\definecolor{lilac}{RGB}{172,114,172}

\DeclareSymbolFont{mymath}{T1}{ybv}{m}{it}
\SetSymbolFont{mymath}{normal}{T1}{ybv}{m}{it}
\DeclareSymbolFontAlphabet{\mathnormal}{mymath}
\DeclareMathSymbol{\vel}{\mathalpha}{mymath}{`v}

\title[Evolution of CNO isotopes]{The evolution of CNO isotopes: the impact of 
  massive stellar rotators}

\author[D.~Romano et al.]{Donatella~Romano,$^{\! 1}$\thanks{E-mail: 
    donatella.romano@inaf.it} Francesca~Matteucci,$^{\! 2, 3, 4}$ 
  Zhi-Yu~Zhang,$^{\! 5, 6}$ Rob~J.~Ivison$^{\, 5, 6}$
  \newauthor
  and Paolo~Ventura$^{\, 7}$\\
  $^{1}$ INAF, Osservatorio di Astrofisica e Scienza dello Spazio, Via Gobetti 
  93/3, I-40129 Bologna, Italy\\
  $^{2}$ Dipartimento di Fisica, Sezione di Astronomia, Universit{\`a} di 
  Trieste, Via Tiepolo 11, I-34131 Trieste, Italy\\
  $^{3}$ INAF, Osservatorio Astronomico di Trieste, Via Tiepolo 11, I-34131 
  Trieste, Italy\\
  $^{4}$ INFN, Sezione di Trieste, Via Valerio 2, I-34127 Trieste, Italy\\
  $^{5}$ Institute for Astronomy, University of Edinburgh, Royal Observatory, 
  Blackford Hill, Edinburgh, EH9 3HJ, UK\\
  $^{6}$ European Southern Observatory, Karl-Schwarzschild-Strasse 2, D-85748, 
  Garching bei M\"unchen, Germany\\
  $^{7}$ INAF, Osservatorio Astronomico di Roma, Via Frascati 33, I-00040 Monte 
  Porzio Catone, Roma, Italy}

\begin{document}

\date{Accepted . Received ; in original form 2019 July 19}

\pagerange{\pageref{firstpage}--\pageref{lastpage}} \pubyear{2019}

\maketitle

\label{firstpage}


\begin{abstract}
Chemical abundances and abundance ratios measured in galaxies provide precious 
information about the mechanisms, modes and time scales of the assembly of 
cosmic structures. Yet, the nucleogenesis and chemical evolution of elements 
heavier than helium are dictated mostly by the physics of the stars and the 
shape of the stellar mass spectrum. In particular, estimates of CNO isotopic 
abundances in the hot, dusty media of high-redshift starburst galaxies offer a 
unique glimpse into the shape of the stellar initial mass function (IMF) in 
extreme environments that can not be accessed with direct observations (star 
counts). Underlying uncertainties in stellar evolution and nucleosynthesis 
theory, however, may hurt our chances of getting a firm grasp of the IMF in 
these galaxies. In this work, we adopt new yields for massive stars, covering 
different initial rotational velocities. First, we implement the new yield set 
in a well-tested chemical evolution model for the Milky Way. The calibrated 
model is then adapted to the specific case of a prototype submillimeter galaxy 
(SMG). We show that, if the formation of fast-rotating stars is favoured in the 
turbulent medium of violently star-forming galaxies irrespective of 
metallicity, the IMF needs to be skewed towards high-mass stars in order to 
explain the CNO isotope ratios observed in SMGs. If, instead, stellar rotation 
becomes negligible beyond a given metallicity threshold, as is the case for our 
own Galaxy, there is no need to invoke a top-heavy IMF in starbursts.
\end{abstract}

\begin{keywords}
nuclear reactions, nucleosynthesis, abundances -- galaxies: abundances -- 
galaxies: evolution -- galaxies: ISM -- stars: abundances -- stars: rotation.
\end{keywords}


\section{Introduction}
\label{sec:intro}

After hydrogen and helium, oxygen, carbon and nitrogen are the most abundant 
elements in the universe. Their seven stable isotopes, $^{12}$C, $^{13}$C, 
$^{14}$N, $^{15}$N, $^{16}$O, $^{17}$O, and $^{18}$O, are synthesised in different 
stellar sites through different processes \citep{1957RvMP...29..547B}: 
(i) the main isotopes of carbon and oxygen come from He burning in stars of all 
masses ($^{12}$C) or in massive stars only ($^{16}$O); (ii) the cold CNO cycle, 
that takes place in the H-burning zones of main sequence and giant branch stars 
in the presence of carbon- (and oxygen-)seed nuclei, leads to the production of 
$^{14}$N and, possibly, $^{13}$C and $^{17}$O (at a lower pace); (iii) in the 
external layers of novae and supernovae (SNe), the activation of the hot CNO 
cycle explains the formation of $^{13}$C, $^{15}$N, and $^{17}$O in H-rich zones, 
and (iv) of $^{15}$N and $^{18}$O in He-rich ones.

While $^{12}$C and $^{16}$O are purely primary elements, i.e., they form 
directly from a hydrogen-helium mixture through a succession of nuclear 
burnings, all other isotopes may have a primary and/or secondary nature, 
depending on whether the metal seeds necessary for their formation are produced 
inside the star, or are already present on the zero-age main sequence. Whether 
it is the primary contribution or the secondary one that prevails, it depends 
primarily on the initial mass and metallicity of the star. For instance, 
fast-rotating intermediate- and high-mass stars produce large amounts of 
primary $^{14}$N at very low metallicities, via production channels that lose 
their effectiveness at higher metallicities \citep{2002A&A...381L..25M}. This 
primary N production was hypothesised long ago \citep{1971Ap&SS..14..179T,
1974ApJ...190..605T} and explains observations of [N/O]\footnote{In this paper, 
we adopt standard definitions of elemental abundance ratios. For two elements X 
and Y, [X/Y]~$\equiv \log(N_{\rm X}/N_{\rm Y}) - \log(N_{\rm X}/N_{\rm Y})_\odot$ is 
the ratio of the abundances by number on a logarithmic scale relative to the 
solar reference value, while X/Y refers to the ratio of the abundances by 
mass.} abundance ratios in metal-poor Galactic dwarfs and ionized H\,II 
regions in our own and other galaxies \citep[see][for early work pointing to 
the need for some primary N production at low 
metallicities]{1974ApJ...189..493S,1975ApJ...199..591S,1978MNRAS.185P..77E,
1978ApJ...220..516P,1979A&A....80..155L,1983A&A...123....1B,
1985MNRAS.217..391M}.

In chemical evolution studies, it is customary to use abundance ratios to gain 
insights into the mechanisms that shape galaxies, and their time scales. As 
matter is more and more strongly processed by succeeding generations of stars, 
one might naively expect an increase in the abundances of secondary elements 
relative to those of primary elements. However, elements that are produced via 
primary nucleosynthetic processes in low-mass stars will behave as secondary 
elements from the point of view of chemical evolution, owing to the long 
lifetimes of their progenitors \citep[e.g.][]{1989MNRAS.239..885M}; similarly, 
primary elements originating in the outer layers of massive stars may exhibit a 
pseudo-secondary character, if their production is boosted at higher 
metallicities because of enhanced mass loss \citep{1992A&A...264..105M,
2000ApJ...541..660H}.

With their multifaceted stellar production sites and nucleosynthesis paths, CNO 
isotopes have repeatedly attracted the attention of chemical evolution 
modelers. Following pioneering works by \citeauthor{1975A&A....43...71A} 
\citeyearpar{1975A&A....43...71A,1977ASSL...67..155A}, 
\citet{1976A&A....52....1V} and \citet{1978ApJ...223..557D}, many authors have 
attempted to explain the origin and evolution of (one or more) CNO isotope 
ratios in the framework of increasingly sophisticated models. These, however, 
have mostly dealt with the Milky Way \citep[e.g.][]{1982ApJ...254..699T,
1991A&A...247L..37M,1996A&A...309..760P,2003MNRAS.342..185R,2008A&A...479L...9C,
2011MNRAS.414.3231K}, with a few exceptions \citep{1993A&A...274..730H,
2008MNRAS.390.1710H,2017MNRAS.470..401R}.

Nowadays, although important uncertainties remain, the constant improvement of 
stellar evolution and nucleosynthesis theory on the one hand, and the advent of 
revolutionary instrumentation of unprecedented sensitivity on the other, allow 
an effective usage of specific CNO isotope ratios as sensible constraints on 
general models of galaxy formation and evolution. In 
\citet{2017MNRAS.470..401R}, we have extended a successful Galactic chemical 
evolution (GCE) model, that reproduces satisfactorily well the CNO abundance 
data for the Milky Way, to other galaxies and shown that the predicted 
$^{16}$O/$^{18}$O abundance ratio may decrease by orders of magnitude if the 
assumed stellar initial mass function (IMF) in galaxy-wide starbursts varies 
from a canonical to a top-heavy one; the same IMF changes have only a minor 
effect on the $^{12}$C/$^{13}$C ratio\footnote{\citet{1993A&A...274..730H} 
reached similar conclusions by relying on less refined chemical evolution 
models and stellar yields.}. Building on this, in \citet{2018Natur.558..260Z} 
we read the remarkably uniform $^{13}$C/$^{18}$O~$\simeq$~1 ratio measured by us 
in four strongly lensed submillimeter galaxies (SMGs) observed with the Atacama 
Large Millimeter/submillimeter Array (ALMA) at redshift $z \sim$~2--3 as a 
clear-cut evidence that the IMF is biased towards massive stars in the most 
extreme starburst events in the universe, and propose to use this ratio as a 
powerful diagnostic of the IMF shape in dust-enshrouded galaxies.

Several independent, more or less direct pieces of observational evidence 
\citep{2005MNRAS.364L..23N,2009ApJ...705..112B,2011MNRAS.415.1647G,
2013ApJ...764..155L,2018Sci...359...69S}, as well as theoretical arguments 
\citep[see][among others]{2008ApJ...674..927A,2010ApJ...720..226P,
2011MNRAS.414.1705P,2018A&A...620A..39J}, support our finding that massive star 
formation is favoured in extreme environments, such as dense star-forming 
regions of high temperature and pressure permeated by intense radiation 
fields. However, new grids of stellar yields have been published recently 
\citep{2018ApJS..237...13L}, which might challenge our interpretation of the 
low $^{13}$C/$^{18}$O ratios observed in SMGs. Indeed, even small differences in 
the yields, when weighted by an appropriate IMF, may result in non-negligible 
effects on galactic-scale model predictions 
\citep[e.g.][]{2010A&A...522A..32R}. Therefore, a reappraisal of our models in 
the light of the newly published nucleosynthesis results is urgently needed.

This paper is organized as follows: in Section~\ref{sec:gce}, we briefly 
outline the main assumptions and ingredients of our chemical evolution models, 
with special emphasis on the adopted nucleosynthesis prescriptions. In 
Section~\ref{sec:res}, we present the model results in comparison to up-to-date 
data for the Milky Way and for the SMGs studied by \citet{2018Natur.558..260Z}. 
We discuss our findings in Section~\ref{sec:dis}; the conclusions follow 
(Section~\ref{sec:end}).

\section{Chemical evolution models}
\label{sec:gce}

\subsection{The Milky Way model}

As in \citet{2017MNRAS.470..401R}, we adopt the two-infall chemical evolution 
model for the Milky Way originally developed by \citet{1997ApJ...477..765C,
2001ApJ...554.1044C}, where a detailed description of the basic equations and 
assumptions can be found. In the next paragraphs, we briefly summarize its main 
features, strengths, and limitations.

The model embraces a multi-zone scheme and divides the Galactic disc in several 
concentric annulii 2~kpc wide. It assumes that the inner halo and thick-disc 
components form out of a first episode of accretion of matter of primordial 
chemical composition. In the first $\sim$1~Gyr of evolution, the fast gas 
consumption resulting from the efficient star formation burst that generates 
the most ancient stellar populations makes the gas density quickly drop below a 
critical threshold, under which the star formation stops. The occurrence of 
such a sudden decrease in star formation was first seen in the [Fe/O] versus 
[O/H] data by \citet[][see also \citealp{1998A&A...338..161F,
2011MNRAS.414.2893F}]{1996ASPC...92..307G}. The thin disc forms later, out of a 
second, almost independent infall episode on time scales that range from less 
than 3~Gyr for the inner disc to nearly a Hubble time for the outer regions 
\citep[equation~2 in][]{2000ApJ...539..235R}. This `inside-out' formation of 
the disc ensures the establishment of radial abundance gradients 
\citep{1976MNRAS.176...31L,1989MNRAS.239..885M}. The adopted star formation 
rate is proportional to both the gas and the total surface mass density 
\citep[see][their equation~2]{2001ApJ...554.1044C}. The efficiency of star 
formation is maximum during the halo/thick-disc phase; during the thin-disc 
phase, it dwindles and becomes a function of the distance from the Galactic 
centre \citep[see][and references therein]{2015ApJ...802..129S}. The adopted 
IMF is that of \citet{2002ASPC..285...86K}, with a slope $x = 1.7$ in the 
high-mass domain ($x = 1.35$ for the extrapolated \citealt{1955ApJ...121..161S} 
law), normalized to unity in the 0.1--100~M$_\odot$ stellar mass 
range\footnote{Many authors have argued that a slope steeper than 
\citeauthor{1955ApJ...121..161S}'s has to be preferred for the local field star 
IMF independently from chemical evolution arguments; see, e.g., 
\citet{1986FCPh...11....1S,2014A&A...564A.102C,2015MNRAS.447.3880R,
2017A&A...599A..17M,2018A&A...620A..39J}.}. Hereinafter, this IMF will be 
referred to as the `canonical IMF'.

The two-infall model has proven able to meet the minimal set of observational 
constraints which should be honored by any successful Galactic chemical 
evolution model, namely: the surface densities of gas, stars and total matter 
in the solar vicinity, as well as their radial profiles; the present-day rates 
of star formation, infall, core-collapse and type Ia SNe in the Galaxy; the 
fractional contribution of metal-poor stars to the total stellar mass in the 
solar neighbourhood; the solar abundances; the trends of the abundances of 
different elements with respect to Fe as a function of [Fe/H]; the degree of 
deuterium astration in the solar vicinity; the radial abundance gradients of 
all the major chemical species \citep[see, e.g.,][for a discussion of the model 
results]{2001ApJ...554.1044C,2006MNRAS.369..295R,2010A&A...522A..32R,
2007A&A...462..943C}.

The adopted model, like all other pure chemical evolution models in the 
literature, makes use of simple, heuristic recipes to treat complex processes, 
such as gas accretion, cooling, star formation, and thermal feedback from 
stars. It does not include complex dynamical processes, such as radial stellar 
migration \citep[but see][]{2015ApJ...802..129S}, and does not deal 
self-consistently with chemical inhomogeneities. However, it is extremely 
efficient in terms of computational costs and allows a quick exploration of the 
free parameter space. Therefore, since our main scope here is to test the 
effect of different nucleosynthesis prescriptions on the model predictions, and 
since we are mostly concerned with average behaviours, similarly to other 
authors \citep{2013ARA&A..51..457N,2018MNRAS.476.3432P} we regard the pure 
chemical evolution model as the optimal tool to comply with our requests.

\subsection{The SMG model}


\begin{table*}
\caption{Nucleosynthesis prescriptions for different models.}
\begin{tabular}{@{}cccccccc@{}}
\hline
\multicolumn{2}{c}{Model$^a$} & LIMS & Super-AGB stars & Massive stars & $\vel_{\rm{rot}}$ & Hypernovae & Novae \\
Milky Way & Prototype SMG & & & & (km s$^{-1}$) & & \\
\hline
\textcolor{lightgreen}{MWG-01} & \textcolor{lightgreen}{SMG-01} & Karakas (2010) & Doherty et al.~(2014a,b) & Nomoto et al.~(2013) & 0 & \ding{55} & \ding{55} \\
\textcolor{mediumgreen}{MWG-02} & \textcolor{mediumgreen}{--} & Karakas (2010) & -- & Nomoto et al.~(2013) & 0 & \ding{55} & \ding{55} \\
\textcolor{darkgreen}{MWG-03} & \textcolor{darkgreen}{--} & \multicolumn{2}{c}{Ventura et al. (2013) \& unpublished} & Nomoto et al.~(2013) & 0 & \ding{55} & \ding{55} \\
\textcolor{mintgreen}{MWG-04} & \textcolor{mintgreen}{--} & \multicolumn{2}{c}{Ventura et al. (2013) \& unpublished} & Nomoto et al.~(2013) & 0 & \ding{51} & \ding{55} \\
\textcolor{lightblue}{MWG-05} & \textcolor{lightblue}{--} & \multicolumn{2}{c}{Ventura et al. (2013) \& unpublished} & Limongi \& Chieffi (2018)$^b$ & 300 & \ding{55} & \ding{55} \\
\textcolor{mediumblue}{MWG-06} & \textcolor{mediumblue}{--} & \multicolumn{2}{c}{Ventura et al. (2013) \& unpublished} & Limongi \& Chieffi (2018)$^b$ & 150 & \ding{55} & \ding{55} \\
\textcolor{darkblue}{MWG-07} & \textcolor{darkblue}{--} & \multicolumn{2}{c}{Ventura et al. (2013) \& unpublished} & Limongi \& Chieffi (2018)$^b$ & 0 & \ding{55} & \ding{55} \\
\textcolor{lightorange}{MWG-08} & \textcolor{lightorange}{SMG-08} & Karakas (2010) & Doherty et al.~(2014a,b) & Limongi \& Chieffi (2018)$^b$ & 300 & \ding{55} & \ding{55} \\
\textcolor{darkorange}{MWG-09} & \textcolor{darkorange}{SMG-09} & Karakas (2010) & Doherty et al.~(2014a,b) & Limongi \& Chieffi (2018)$^b$ & 150 & \ding{55} & \ding{55} \\
\textcolor{mediumred}{MWG-10} & \textcolor{mediumred}{--} & Karakas (2010) & Doherty et al.~(2014a,b) & Limongi \& Chieffi (2018)$^b$ & 0 & \ding{55} & \ding{55} \\
\textcolor{jazz}{MWG-11} & \textcolor{jazz}{SMG-11} & \multicolumn{2}{c}{Ventura et al. (2013) \& unpublished} & Limongi \& Chieffi (2018)$^c$ & var$\,^d$ & \ding{55} & \ding{51} \\
\textcolor{lilac}{MWG-12} & \textcolor{lilac}{SMG-12} & Karakas (2010) & Doherty et al.~(2014a,b) & Limongi \& Chieffi (2018)$^c$ & var$\,^d$ & \ding{55} & \ding{51}\\
\hline
\end{tabular}
\label{tab:nuc}
       \begin{flushleft}
       \emph{Notes.} $^a$We identify a particular model with the notation 
       XXX-YY, where XXX individuates the galaxy (MWG: Milky Way galaxy, SMG: 
       submillimeter galaxy) and YY is a number referring to the selected yield 
       set combination. All different nucleosynthesis prescriptions have been 
       tested against the Milky Way data, whilst only a subset of prescriptions 
       has been used in SMG models. $^b$We adopt their recommended yield set 
       (set R). $^c$Their set R is used for [Fe/H]~$< -$1, while for 
       [Fe/H]~$\ge -$1 the mass range for full collapse to black holes is 
       reduced (60--100~M$_\odot$). $^d \vel_{\rm{rot}} = 300$~km s$^{-1}$ for 
       [Fe/H]~$< -$1, $\vel_{\rm{rot}} = 0$ for [Fe/H]~$\ge -$1.
       \end{flushleft}
\end{table*}


We adopt the one-zone model for the typical SMG discussed in 
\citet{2017MNRAS.470..401R} and \citet{2018Natur.558..260Z}. The model relies 
on the evolutionary sequence advocated by \citet{2014ApJ...782...68T}, 
according to which SMGs are the precursors of passively-evolving massive 
elliptical galaxies. Fresh gas is accreted at early times on a short time scale 
at an exponentially decreasing rate,
\begin{equation}
\frac{{\rm d}\mathscr{M}_{\rm inf}}{{\rm d}t} \propto {\rm e}^{-t/\tau},
\end{equation}
where $\mathscr{M}_{\rm{inf}} = 4 \times 10^{11}$~M$_\odot$ is the total baryonic 
mass accreted by the system and $\tau =$ 50~Myr is the infall time scale. The 
gas forms stars following a 
\citeauthor{1998ApJ...498..541K}-\citeauthor{1959ApJ...129..243S} law,
\begin{equation}
\psi(t) = \nu\,\mathscr{M}_{\rm{gas}}^k(t), 
\end{equation}
where $\nu$ is the star formation efficiency, $k = 1$ and 
$\mathscr{M}_{\rm{gas}}(t)$ is the mass of neutral gas available for star 
formation at each time \citep{1959ApJ...129..243S,1998ApJ...498..541K}. The 
star formation is halted when a stellar mass $\mathscr{M}_{\rm{stars}} \simeq 2 
\times 10^{11}$~M$_\odot$ is attained; a strong galactic wind triggered by SN 
explosions and AGN activity is supposed to clean the galaxy of its residual gas 
at this point. We consider a rather continuous or a bursty star formation 
regime, as well as a canonical or a top-heavy IMF ($x = 1.1$ in the high-mass 
domain), and vary the duration of the star formation episode(s). We compare the 
model outputs in the different cases.

\subsection{Nucleosynthesis prescriptions}

In our computations the instantaneous recycling approximation is relaxed, i.e. 
we take into account in detail the finite stellar lifetimes. This is necessary 
in order to treat properly elements that are produced on different time scales 
by stars of different initial masses and various chemical compositions. The 
adopted nucleosynthesis prescriptions are summarized in Table~\ref{tab:nuc} and 
briefly discussed in the remainder of this section; a more thorough 
description can be found in the original papers.

\subsubsection{Single stars}

The stellar yields for low- and intermediate-mass stars (LIMS; 1~$\lesssim 
m/{\rm M}_{\odot} \lesssim$~6) are either from \citet{2010MNRAS.403.1413K} or 
from \citet{2013MNRAS.431.3642V}. The latter authors also provide detailed 
yields for super-asymptotic giant branch (AGB) stars 
(6~$\lesssim m/{\rm M}_{\odot} \lesssim$~8--9) and (unpublished) yields for 
super-solar metallicity stars; both ingredients are often missing from GCE 
calculations. The nucleosynthetic yields for super-AGB stars computed by 
\citet{2014MNRAS.437..195D,2014MNRAS.441..582D} are used to complement the LIMS 
grid of \citet{2010MNRAS.403.1413K}, apart from one model where this super-AGB 
contribution is set to zero (model MWG-02). Although the stellar mass range 
covered by super-AGB stars is quite modest, they are found to contribute 
non-negligible amounts of $^{14}$N on a galactic scale (see 
Section~\ref{sec:res}).

For massive stars, we adopt the grid of yields suggested by 
\citet{2013ARA&A..51..457N}, largely based on work published in 
\citet{2006ApJ...653.1145K,2011MNRAS.414.3231K} and extending to super-solar 
metallicities (models from MWG-01 to MWG-04, and model SMG-01). These yields 
account for some important physical processes that impact deeply the 
nucleosynthetic outcome of a star, such as mass loss, the `mixing and fallback' 
process \citep[][and references therein]{2002ApJ...565..385U}, and the 
occurrence of highly energetic explosions \citep[the so-called hypernovae, 
according to the terminology first introduced by][]{1998ApJ...494L..45P}. In 
particular, all stars above 20~M$_\odot$ are assumed to explode as hypernovae in 
model MWG-04, while they end up as ordinary core-collapse SNe in the other 
models. The important effects of stellar rotation, however, are not accounted 
for in the \citeauthor{2013ARA&A..51..457N}'s grid. Therefore, we also 
implement in our GCE code \citeauthor{2018ApJS..237...13L}'s 
\citeyearpar{2018ApJS..237...13L} recommended yield set (set~R), where: (i) 
mass loss and (ii) rotation are taken into account; (iii) stars with initial 
mass in the range 13--25~M$_\odot$ eject 0.07~M$_\odot$ of $^{56}$Ni after going 
through mixing and fallback in the inner SN regions; (iv) more massive objects 
fully collapse to black holes, which results in a null injection of Fe from 
$m >$~30~M$_\odot$ stars\footnote{In models MWG-11 and MWG-12, the mass range 
for full collapse to black holes is reduced to 60--100~M$_\odot$ for 
[Fe/H]~$\ge -$1.}.

All the adopted yields are dependent on the initial mass and metallicity of the 
stars. Since they are necessarily computed for a limited number of points in 
the ($m$, $Z$) space, some interpolation in mass and metallicity has to be 
performed between published adjacent values. In particular, although we 
implement in the code up-to-date nucleosynthesis prescriptions for super-AGB 
stars, a most uncertain interpolation of the yields is still required in the 
mass range pertaining to low-mass core-collapse SNe ($\sim$9--12~M$_\odot$). On 
top of that, when adopting the \citet{2013ARA&A..51..457N} yield sets for 
metallicities $Z \neq 0$, extrapolation is needed from 40 to 100~M$_\odot$. 
Possible spurious effects can be introduced as artifacts of the 
interpolation/extrapolation procedures, and one should keep this in mind when 
comparing GCE model predictions to observations. Denser grids of stellar yields 
are the only way to overcome this problem.

\subsubsection{Binary systems}
\label{sec:bs}

The nucleosynthetic outcome of binary systems ending their lives as type Ia SNe 
is included in our models: we adopt the single-degenerate scenario \citep[][and 
references therein]{2001ApJ...558..351M} for the progenitors and 
nucleosynthetic yields from \citet{1999ApJS..125..439I}.

While type Ia SNe contribute negligible amounts of CNO nuclei to the 
interstellar medium, classical nova explosions could significantly impact the 
evolution of the abundances of $^{13}$C, $^{15}$N and $^{17}$O on a Galactic 
scale \citep[see][for a review]{2007JPhG...34..431J}.

The nova contribution to the synthesis of $^{13}$C, $^{15}$N and $^{17}$O is 
included in two of our Milky Way models (models MWG-11 and MWG-12) by assuming 
average yields per nova system, similarly to what is done in \citet[][see next 
paragraphs]{2017MNRAS.470..401R}. Models~SMG-11 and SMG-12, that formally share 
the same nucleosynthesis prescritions with, respectively, model~MWG-11 and 
model~MWG-12, do not include a nova contribution in practice: in fact, before 
a newborn nova system can give rise to strong enough outbursts, it is necessary 
to wait at least 1~Gyr to ensure that the white dwarf has sufficiently cooled 
down (see below) and no SMG is expected to be forming stars on time scales 
longer than this. Therefore, there is basically no room for any CNO isotope 
pollution from novae in SMGs.

In models~MWG-11, MWG-12, SMG-11 and SMG-12 nova nucleosynthesis is implemented 
following the prescriptions of \citet[][and references 
therein]{1999A&A...352..117R,2017MNRAS.470..401R}, as briefly recalled 
hereunder; the interested reader is referred to the original papers for more 
details:
\begin{description}
\item The rate of formation of nova systems at a given time $t$ is computed as 
  a fraction $\alpha$ of the white dwarf birth rate at a previous time 
  $t - \Delta t$, where $\Delta t =$ 1~Gyr is a suitable average time delay 
  that guarantees that the white dwarfs cool down to temperatures low enough to 
  ensure strong outbursts. The free parameter $\alpha$ is assumed to be 
  constant in time \citep[but see][]{2019arXiv190609130G} and set to reproduce 
  the current nova outburst rate in the Galaxy.
\item Each nova system is assumed to experience 10$^4$ powerful eruptions on an 
  average \citep{1978MNRAS.183..515B}.
\item The average masses ejected in the form of $^{13}$C, $^{15}$N and $^{17}$O 
  in each outburst are fixed by the request that the observed declining trends 
  of $^{12}$C/$^{13}$C, $^{14}$N/$^{15}$N and $^{18}$O/$^{17}$O in the last 4.5~Gyr 
  in the solar vicinity are reproduced. For model~MWG-11 (MWG-12), these masses 
  are: 
  $M^{\mathrm{ejec}}_{\mathrm{^{13}C}} = 4$ $(2) \times 10^{-7}$~M$_\odot$, 
  $M^{\mathrm{ejec}}_{\mathrm{^{15}N}} = 5$ $(6.5) \times 10^{-8}$~M$_\odot$ and 
  $M^{\mathrm{ejec}}_{\mathrm{^{17}O}} = 3$ $(1.5) \times 10^{-8}$~M$_\odot$. These 
  quantities are in reasonable agreement with those emerging from hydrodynamic 
  nova models, though a high variability is found in the theoretical nova 
  yields in dependence of the parameters adopted to produce the outburst 
  \citep[see figure~6 in][and the relevant discussion 
    therein]{2017MNRAS.470..401R}.
\end{description}

\section{Results}
\label{sec:res}

In the following, we first present our GCE model results in comparison with 
up-to-date CNO measurements in the Milky Way and select the stellar yields that 
guarantee the best fit to the Galactic data (Section~\ref{sec:mw}). In the 
second place, we analyse a reduced set of models in comparison to CNO data for 
a sample of SMGs caught at the peak of their star formation activity and make 
some considerations about the shape of the IMF in these extreme starbursts 
(Section~\ref{sec:smg}).


\begin{figure*}
\begin{center}
\hspace{1.1cm}
\includegraphics[width=15.92cm]{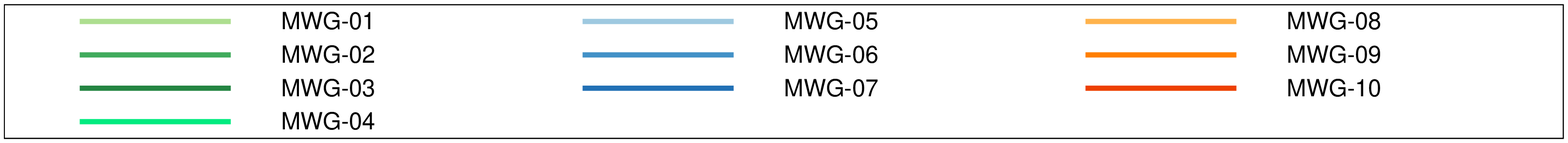}

\includegraphics[width=7.8cm]{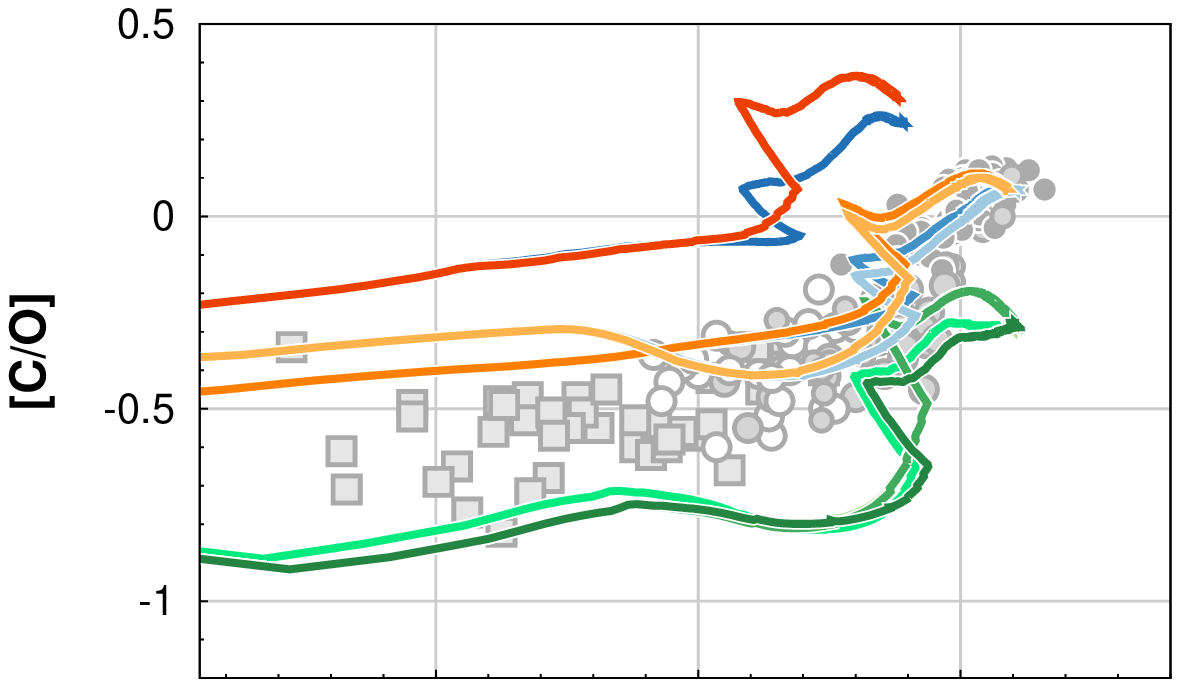}
\hspace{1cm}
\includegraphics[width=7.8cm]{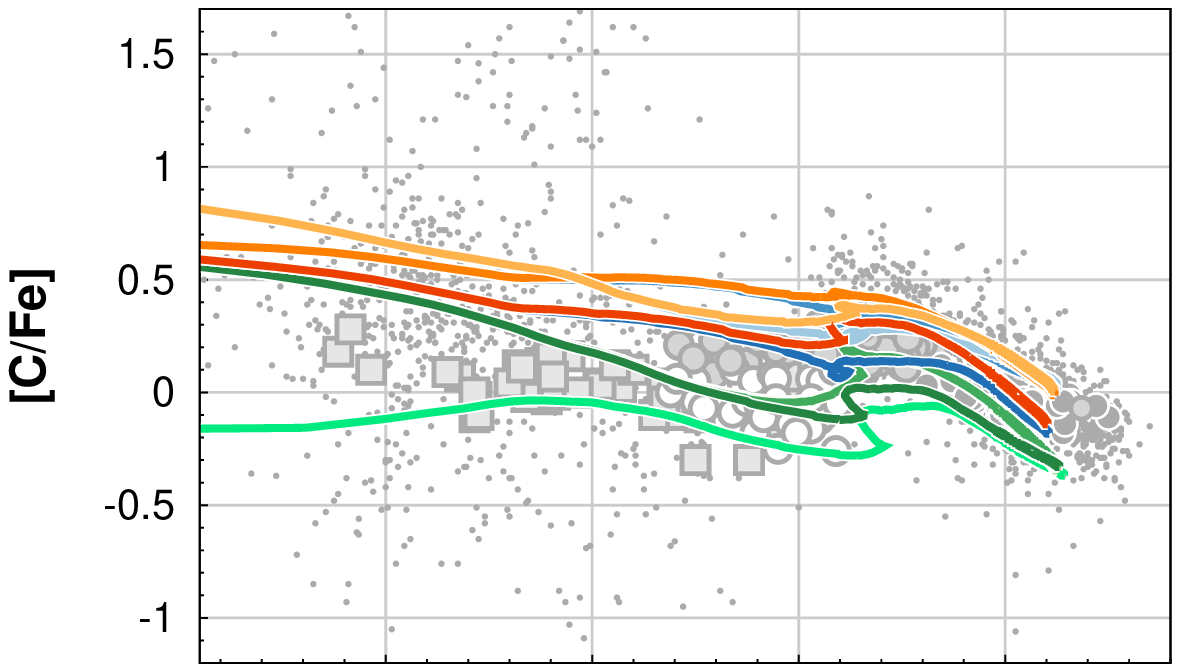}

\vspace{-0.35cm}

\includegraphics[width=7.8cm]{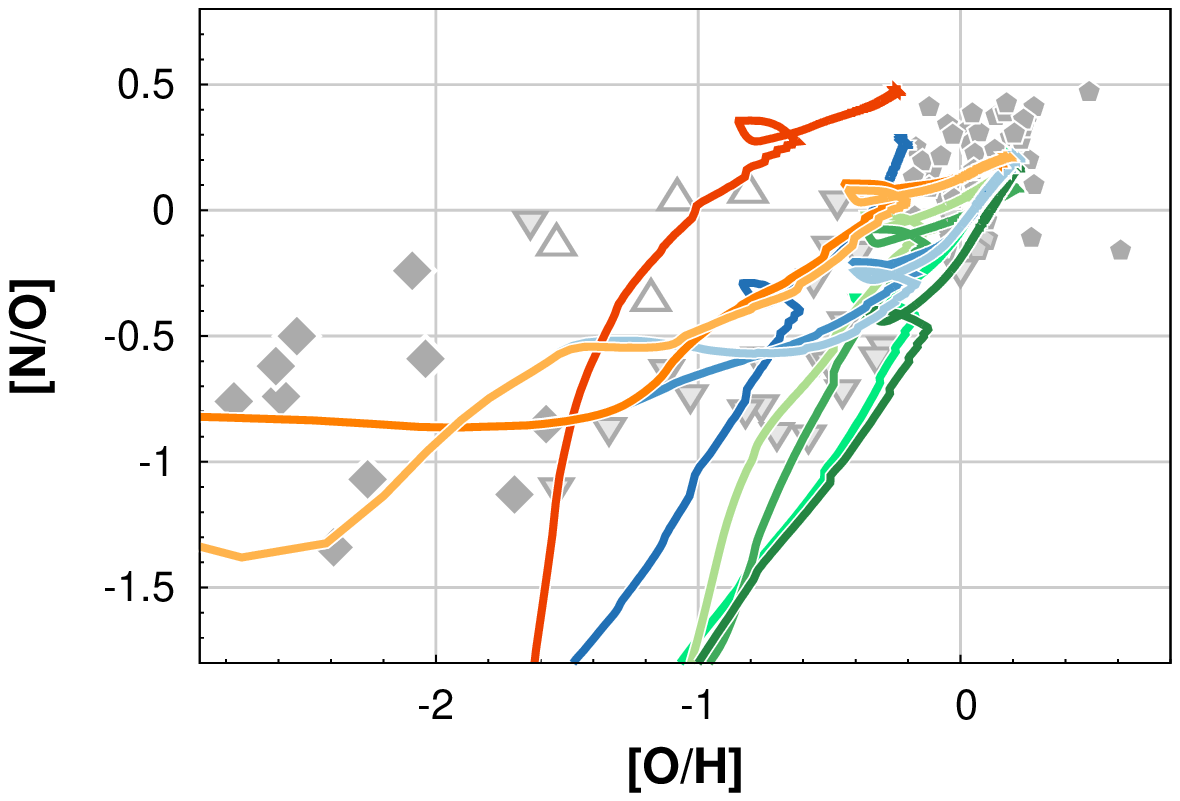}
\hspace{1cm}
\includegraphics[width=7.8cm]{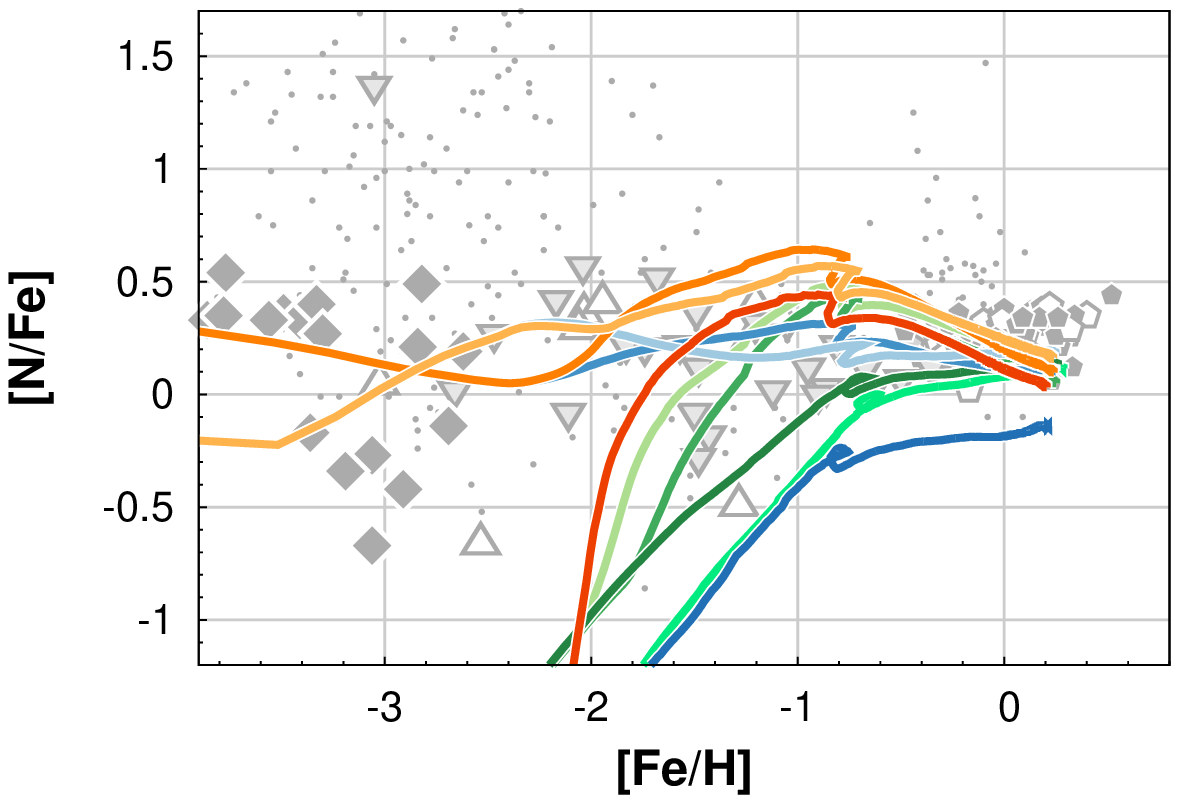}

\caption{ \emph{Left-hand panels:} [C/O]--[O/H] (top) and [N/O]--[O/H] (bottom) 
diagrams for solar neighbourhood stars. Data for carbon are from 
\citet[][\emph{small dark-grey circles:} thin-disc stars; \emph{small 
light-grey circles:} thick-disc stars; \emph{medium-sized light-grey circles:} 
high-$\alpha$ halo stars; \emph{medium-sized empty circles:} low-$\alpha$ halo 
stars]{2014A&A...568A..25N} and \citet[][\emph{light-grey 
squares}]{2019A&A...622L...4A}; data for nitrogen are from 
\citet[][\emph{upside-down triangles}]{2004A&A...421..649I}, 
\citet[][\emph{diamonds}]{2005A&A...430..655S}, \citet[][\emph{empty 
triangles}]{2014AJ....147..136R} and \citet[][\emph{small dark-grey pentagons:} 
thin-disc stars; \emph{small light-grey pentagons:} thick-disc and halo stars; 
\emph{medium-sized empty pentagons:} stars that could not be ascribed to a 
specific Galactic component]{2016A&A...591A..69S}. \emph{Right-hand panels:} 
[C/Fe]--[Fe/H] (top) and [N/Fe]--[Fe/H] (bottom) diagrams for solar 
neighbourhood stars. Data as left-hand panels; also plotted are data from the 
SAGA database and Hypatia catalog \citep[][respectively, 
\emph{dots}]{2008PASJ...60.1159S,2014AJ....148...54H}. See the text for a 
discussion of the data. In all panels, the model predictions 
\emph{(solid lines)} are color-coded according to the adopted nucleosynthesis 
prescriptions (see Table~\ref{tab:nuc}). In this and the following figure 
observed and theoretical abundance ratios are normalized to solar values from 
\citet{2009ARA&A..47..481A}.}
\label{fig:CNO}
\end{center}
\end{figure*}



\begin{figure}
\begin{center}
\includegraphics[width=7.8cm]{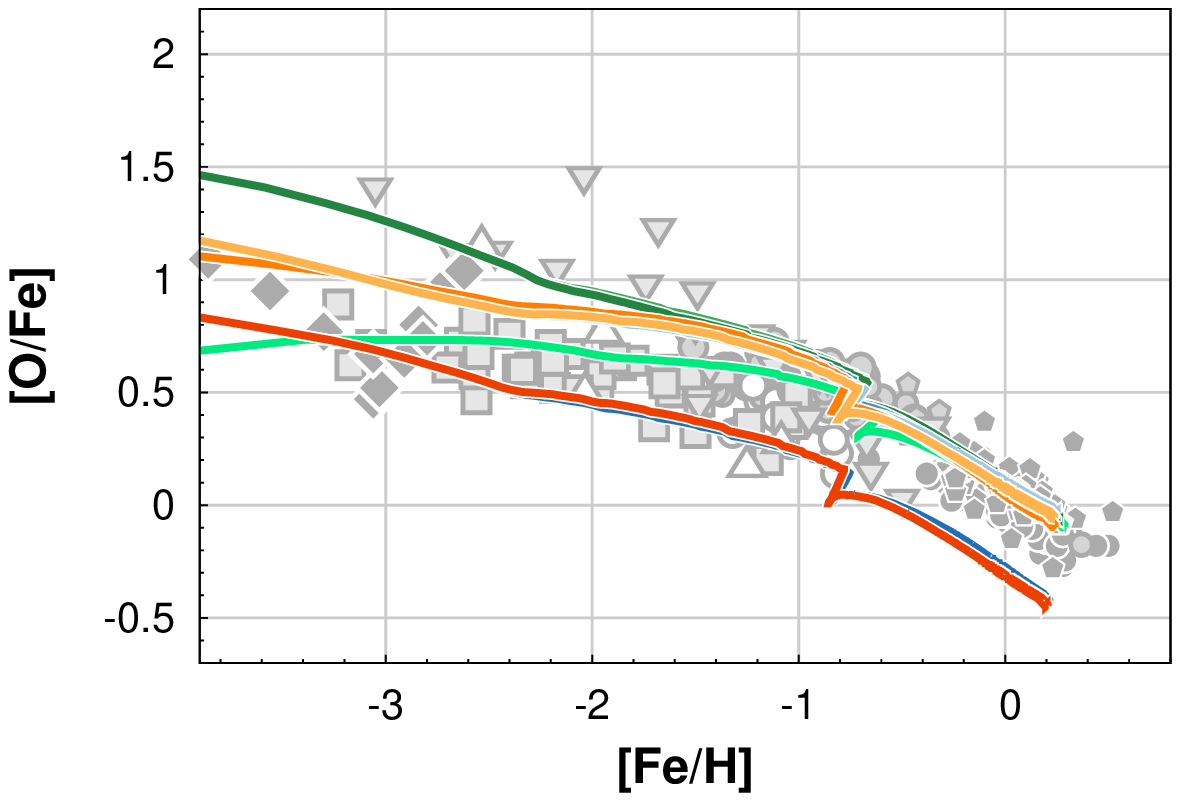}
\caption{ [O/Fe]--[Fe/H] diagram for solar neighbourhood stars. Data are from 
\citet[][\emph{upside-down triangles}]{2004A&A...421..649I}, 
\citet[][\emph{diamonds}]{2005A&A...430..655S}, \citet[][\emph{small dark-grey 
circles:} thin-disc stars; \emph{small light-grey circles:} thick-disc stars; 
\emph{medium-sized light-grey circles:} high-$\alpha$ halo stars; 
\emph{medium-sized empty circles:} low-$\alpha$ halo 
stars]{2014A&A...568A..25N}, \citet[][\emph{empty 
triangles}]{2014AJ....147..136R}, \citet[][\emph{small dark-grey pentagons:} 
thin-disc stars; \emph{small light-grey pentagons:} thick-disc and halo stars; 
\emph{medium-sized empty pentagons:} stars that could not be ascribed to a 
specific Galactic component]{2015A&A...576A..89B} and \citet[][\emph{light-grey 
squares}]{2019A&A...622L...4A}. The model predictions \emph{(solid lines)} are 
color-coded according to the adopted nucleosynthesis prescriptions (see 
Table~\ref{tab:nuc} and explanatory legend in Fig.~\ref{fig:CNO}).}
\label{fig:O}
\end{center}
\end{figure}


\subsection{Milky Way galaxy}
\label{sec:mw}

\subsubsection{Abundance ratios of solar neighbourhood stars}
\label{sec:aburat}

Low-mass stars have lifetimes comparable to the age of the universe. Therefore, 
they are fossils bearing the imprints of the chemical enrichment processes that 
shaped their host galaxies from the very beginning.

It is well documented, though, that small mass field stars climbing the upper 
red giant branch undergo some important abundance changes. The original 
composition of their outer layers is modified markedly: the abundance of 
$^{14}$N increases, at the expenses of that of $^{12}$C, and the $^{12}$C/$^{13}$C 
ratio decreases to values close to, though higher than, the equilibrium value 
\citep[e.g.][]{2000A&A...354..169G}. Unluckily, limitations imposed by current 
observational capabilities prevent any firm conclusion about the oxygen isotope 
ratios \citep[e.g.][]{2015A&A...578A..33L}. Unusually high C abundances in some 
cases may be attributable to mass transfer from a post-AGB companion across a 
binary system via Roche lobe overflow \citep[][and references 
therein]{2005ApJ...625..825L}. Last but not least, some low-metallicity stars 
currently found in the solar neighbourhood have likely been accreted from dwarf 
satellites and, thus, reflect distinct evolutionary paths. If observed stellar 
samples are not cleaned off for these effects, the comparison between GCE model 
predictions and observations via widely-used diagnostic diagrams, such as 
[C/Fe] versus [Fe/H] or [N/O] versus [O/H], can easily lead to 
misinterpretation. It is, thus, advisable to rely on statistically significant 
samples of single, Sun-like stars, formed \emph{in situ} and, possibly, 
analysed homogeneously.

\paragraph{[X/Fe] versus [Fe/H] plots.}

The grey dots in Fig.~\ref{fig:CNO}, right-hand panels, show the [C/Fe] and 
[N/Fe] abundance ratios (upper and lower panel, respectively) of solar 
neighbourhood stars in the SAGA database \citep{2008PASJ...60.1159S} and 
Hypatia catalog \citep{2014AJ....148...54H}. For illustration purposes, we plot 
stars in different evolutive stages, as well as binary stars: as expected, a 
large dispersion is seen in the data at all metallicities. No attempts are made 
to homogenize the data, that come from different sources, except for a 
rescaling to the same solar abundances (taken from 
\citealt{2009ARA&A..47..481A} in this work).

Symbols in different shades of grey represent measurements for unevolved stars 
from \citet[][upside-down triangles]{2004A&A...421..649I}, 
\citet[][diamonds]{2005A&A...430..655S}, 
\citet[][circles]{2014A&A...568A..25N}, 
\citet[][triangles]{2014AJ....147..136R}, 
\citet[][pentagons]{2016A&A...591A..69S} and 
\citet[][squares]{2019A&A...622L...4A}.

\citet{2014A&A...568A..25N} have determined precise C (and O) abundances for 
151~F and G main-sequence stars with metallicities in the range 
$-$1.6~$<$~[Fe/H]~$<$~0.5, classified into four different populations: 
thin-disc, thick-disc, high-$\alpha$ halo and low-$\alpha$ halo stars (shown, 
respectively, as small dark-grey, small light-grey, medium-sized light-grey and 
empty circles in Fig.~\ref{fig:CNO}, upper panels). These authors have used 
plane parallel (1D) model atmospheres for their analysis and applied non-LTE 
corrections to the abundances derived from the C\,I 
$\lambda \lambda$5052,5380 lines and $\lambda$7774~O\,I triplet. As seen 
from Fig.~\ref{fig:CNO}, upper right-hand panel, these precise abundances 
define a tight trend in the [C/Fe]--[Fe/H] plane. Under the reasonable 
assumption \citep[][and references therein]{2014A&A...568A..25N} that the 
low-$\alpha$ halo stars formed in dwarf galaxies that were later accreted by 
the Milky Way and, thus, do not trace the local enrichment history, the 
observed trend for [Fe/H]~$> -$1.5 is fitted at best by models~MWG-07 and 
MWG-10, that assume the yields by \cite{2018ApJS..237...13L} for non-rotating 
massive stars (see Table~\ref{tab:nuc}). Marginal agreement is obtained by the 
models that adopt the yields by \cite{2018ApJS..237...13L} for rotating massive 
stars (models~MWG-05, MWG-06, MWG-08, MWG-09) and by models~MWG-01 and MWG-02, 
that assume the yields by \citet{2013ARA&A..51..457N} for (non-rotating) 
massive stars without hypernovae, as well as the yields by 
\citet{2010MNRAS.403.1413K} for LIMS (C production from super-AGB stars as 
issued in \citealt{2014MNRAS.437..195D,2014MNRAS.441..582D} turns out to be 
negligible). 

The behaviour of C (and O) abundances at lower metallicities has been 
reassessed recently by \citet{2019A&A...622L...4A}. Basing on 3D, non-LTE 
results for 39 turn-off stars with [Fe/H]~$\le -$1.5, they find that [C/Fe] is 
almost solar and stays nearly flat with metallicity. Interestingly, the 
abundances of three stars previously identified as carbon-enhanced stars with 
[C/Fe] in excess of 1 are significantly revised downwards, to [C/Fe]~$\simeq$ 
0.1--0.3 \citep[see][their section~4, for a thorough 
discussion]{2019A&A...622L...4A}. When compared with our model predictions, 
these new measurements seem to point to the need for some hypernova pollution 
at low metallicities (cfr. the predictions from model MWG-04, in which all 
stars above 20~M$_\odot$ end up with a hypernova explosion, with those from 
model MWG-03, which is the same as model MWG-04, but with all massive stars 
exploding as ordinary SNe; Fig.~\ref{fig:CNO}, upper right-hand panel).

We now examine the behaviour of [N/Fe] versus [Fe/H]. An inspection of 
Fig.~\ref{fig:CNO}, lower right-hand panel, immediately shows that a 
significant dispersion characterizes the data at low-metallicities, even when 
only unevolved stars are considered \citep{2004A&A...421..649I,
2005A&A...430..655S,2014AJ....147..136R}. Below [Fe/H]~$\sim -$1.5, some 
observations can be explained only by invoking a significant primary N 
production from massive stars \citep[a well-known result; see previous work 
by][]{1986MNRAS.221..911M,2006A&A...449L..27C,2010A&A...522A..32R,
2018MNRAS.476.3432P}. However, a few low [N/Fe] values point to a milder N 
enrichment. The N yields of massive stars depend critically on the stellar 
rotational velocity, therefore, a large dispersion in the N abundances would 
naturally arise during the Galactic halo assembly, because of the highly 
inhomogeneous evolution: regions polluted by non-rotating massive stars will 
display a N content significantly lower than regions where matter was processed 
through one or more fast rotators. Assessing the evolution of the CNO elements 
in a more realistic, inhomogeneous medium would be extremely valuable, and can 
be done by means of detailed hydrodynamical simulations. A quantitative 
explanation of the observed scatter, however, is beyond the scope of the 
present work and will be addressed in a future paper. It suffices to note here 
that the dispersion is sensibly reduced at disc metallicities, where reliable N 
abundances can be derived for statistically significant samples of dwarf stars 
from the $\lambda$3360 NH molecular band \citep{2016A&A...591A..69S}. Overall, 
we find that models MWG-05 and MWG-06, with yields from 
\cite{2018ApJS..237...13L} for rotating massive stars and from 
\citet{2013MNRAS.431.3642V} for LIMS and super-AGB stars, provide an excellent 
fit to the average [N/Fe] ratios over the full range of metallicities.

The run of [O/Fe] as a function of [Fe/H] is shown in Fig.~\ref{fig:O}. All 
models provide an adequate fit to the observed trend, apart from models MWG-07 
and MWG-10, that are computed with the yields from non-rotating massive stars 
by \citet[][their set~R]{2018ApJS..237...13L}, which severely underestimate the 
[O/Fe] ratios of disc stars. If the 3D oxygen abundances by 
\cite{2019A&A...622L...4A} have to be given a higher weight, it is possible 
that in the early Galaxy a large fraction of high-mass stars exploded as 
hypernovae, with an energy release substantially higher than that of standard 
SNe (cfr. the predictions from model MWG-04, in which all stars above 
20~M$_\odot$ end up as hypernovae, with those from model MWG-03, in which the 
same stars explode as normal core-collapse SNe).

\paragraph{[X/O] versus [O/H] plots.}

The [C/O]--[O/H] and [N/O]--[O/H] diagrams shown in Fig.~\ref{fig:CNO}, 
left-hand panels, are particularly useful. At low metallicities, they trace the 
enrichment from massive stars with minimum linkage to one of the most uncertain 
parameters of stellar evolution, namely, the location of the mass cut (in most 
extant one- and two-dimensional models of the latest stages of massive star 
evolution, this is the boundary between the collapsed core and the ejected 
mantle); by contrast, the choice of the mass cut influences considerably the 
theoretical trends in the [X/Fe] versus [Fe/H] planes. At disc metallicities, 
the rise in [C/O] and [N/O] can be related to the late contribution to C and N 
production from low-metallicity, low-mass stars, and directly compared to the 
release of C and N from relatively more metal-rich massive stars, which occurs 
on much shorter time scales.

Moreover, from a theoretical point of view, the trends with metallicity as 
gauged by oxygen abundance are free from the uncertainties that arise from the 
inclusion of type Ia SNe in the models (these systems produce a large fraction 
of the solar iron, but negligible amounts of oxygen). However, from a 
spectroscopist's standpoint, the abundance estimates are less robust for O than 
for Fe. Oxygen abundances are affected by non-negligible non-LTE corrections 
when derived from the O\,I triplet, and by 3D effects (as well as a 
Ni\,I blend) when derived from the forbidden [O\,I] line at 
6300~\AA; the latter line is also too weak to be detectable in the majority of 
halo stars \citep[see][and references therein]{2014A&A...568A..25N}. 
Furthermore, the O abundances of metal-poor stars derived from near-UV OH lines 
\citep[such as those used by][]{2004A&A...421..649I} are found to be 
sistematically higher than the ones obtained from other indicators (see 
Fig.~\ref{fig:O}).


\begin{figure*}
\begin{center}
\includegraphics[width=7.8cm]{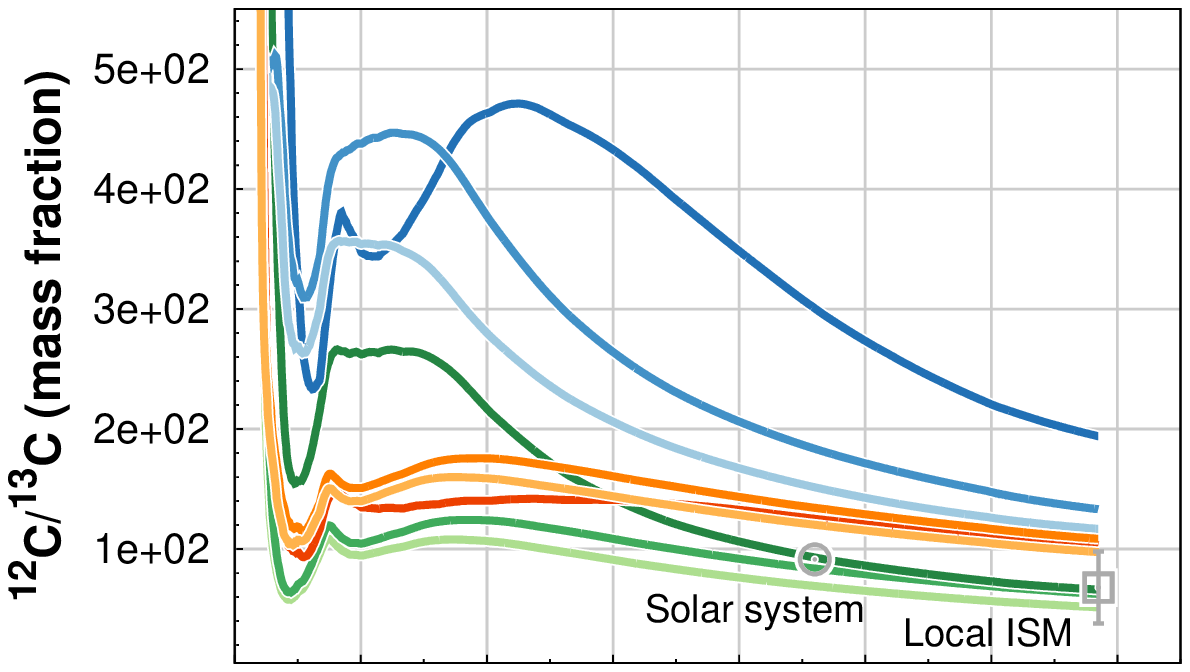}
\includegraphics[width=6.435cm]{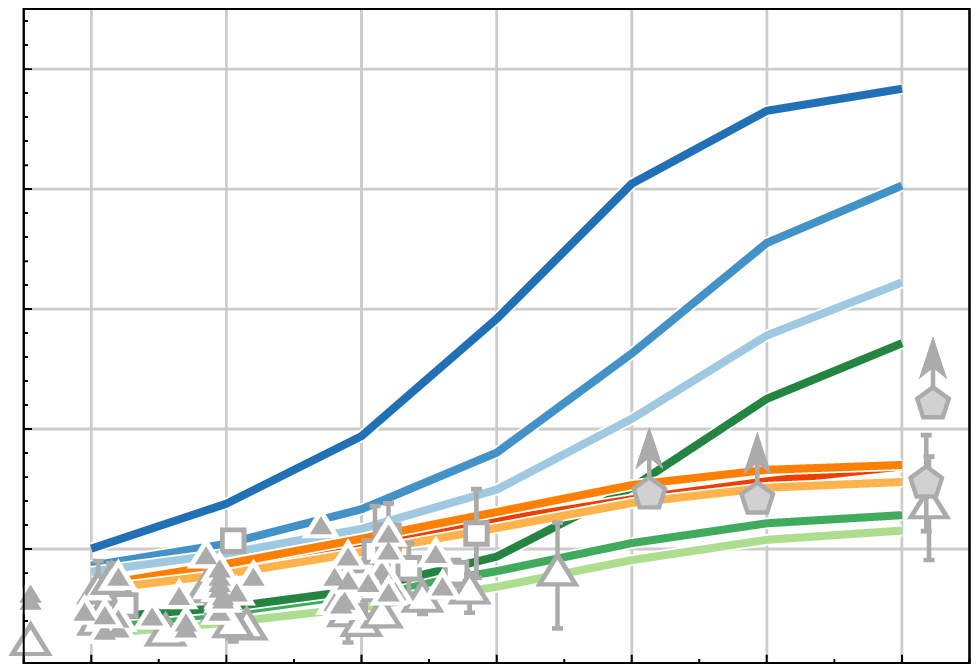}

\vspace{-0.35cm}

\includegraphics[width=7.8cm]{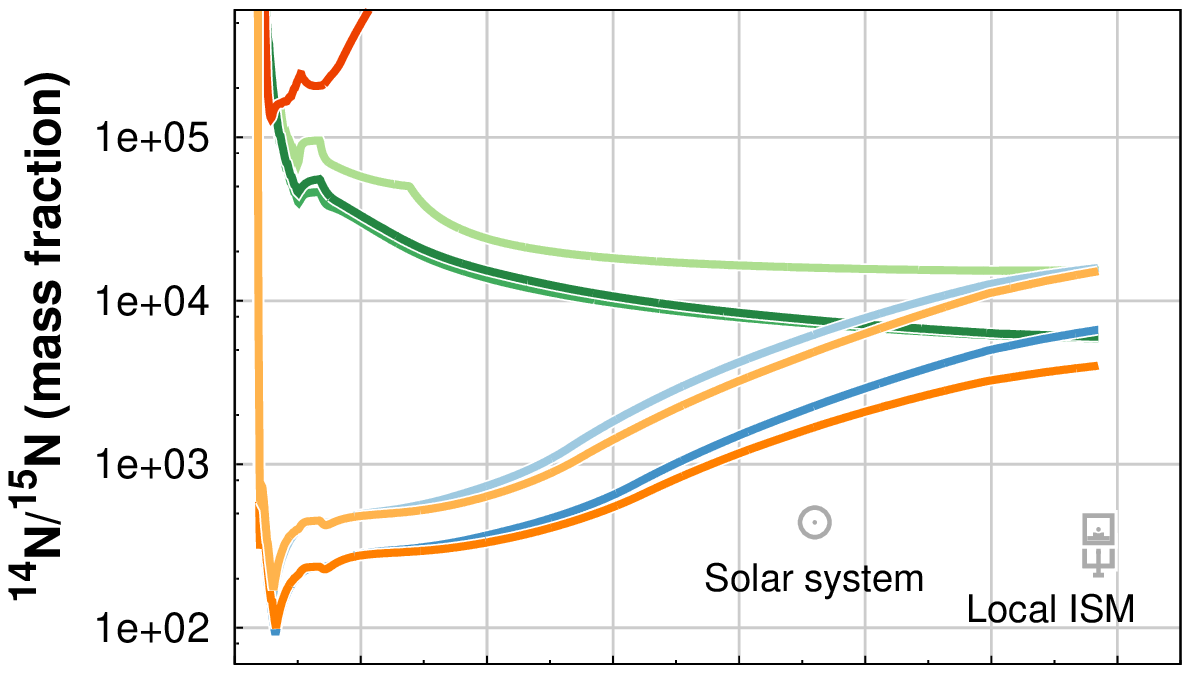}
\includegraphics[width=6.435cm]{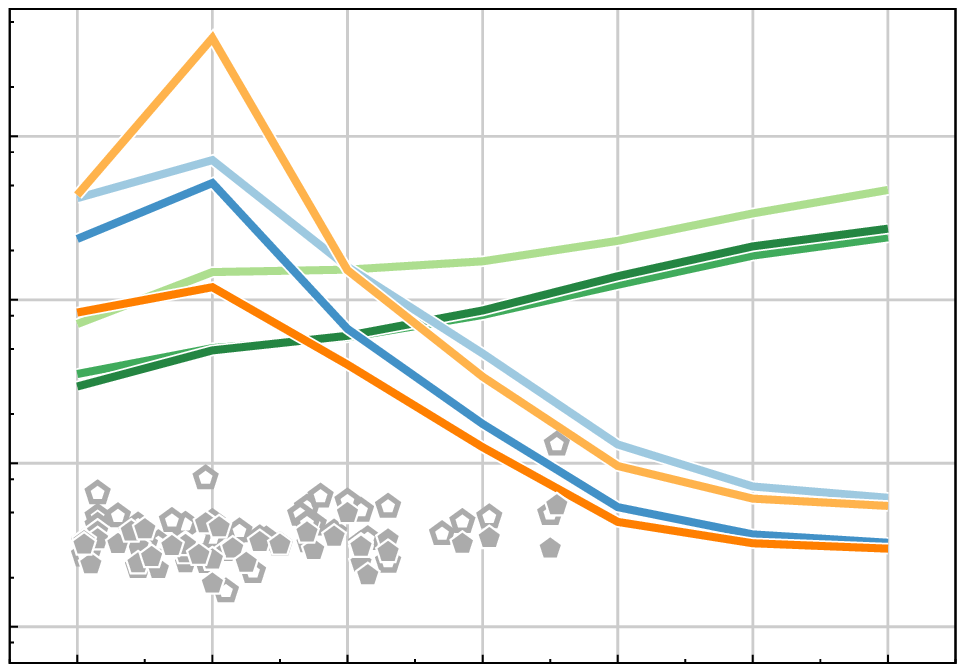}

\vspace{-0.35cm}

\includegraphics[width=7.8cm]{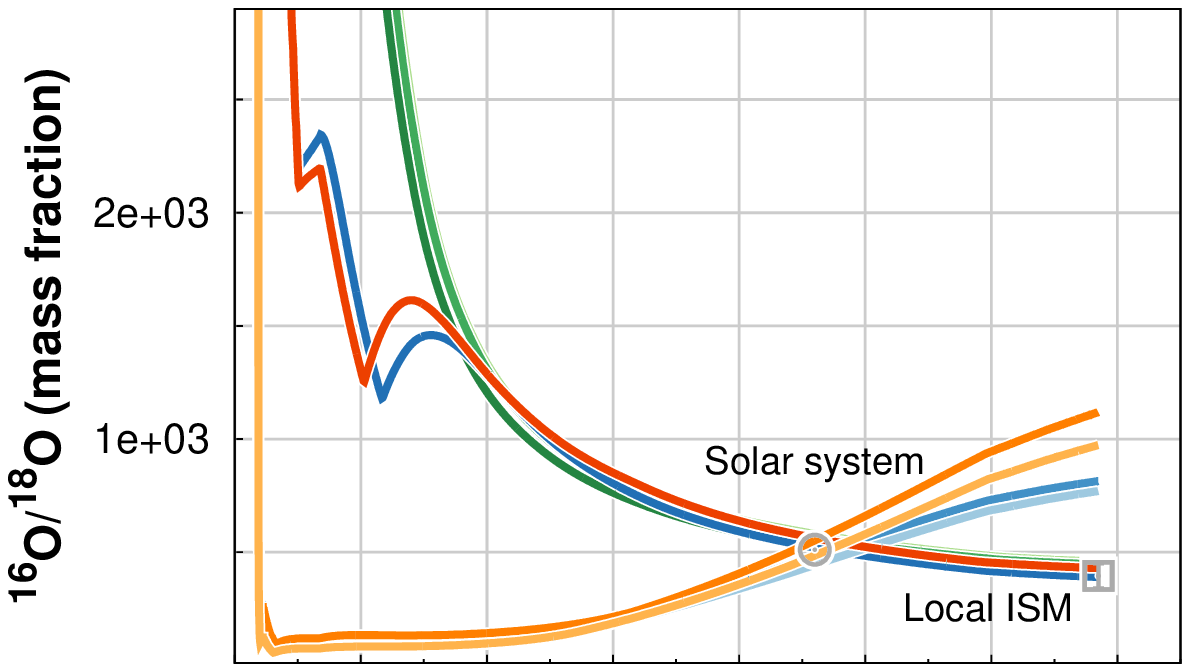}
\includegraphics[width=6.435cm]{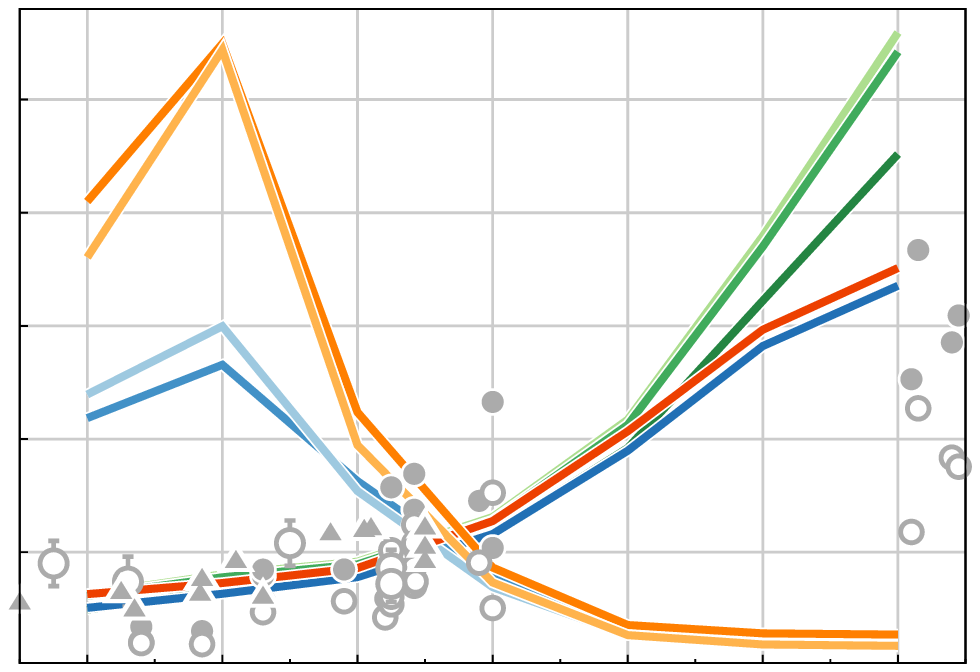}

\vspace{-0.35cm}

\includegraphics[width=7.8cm]{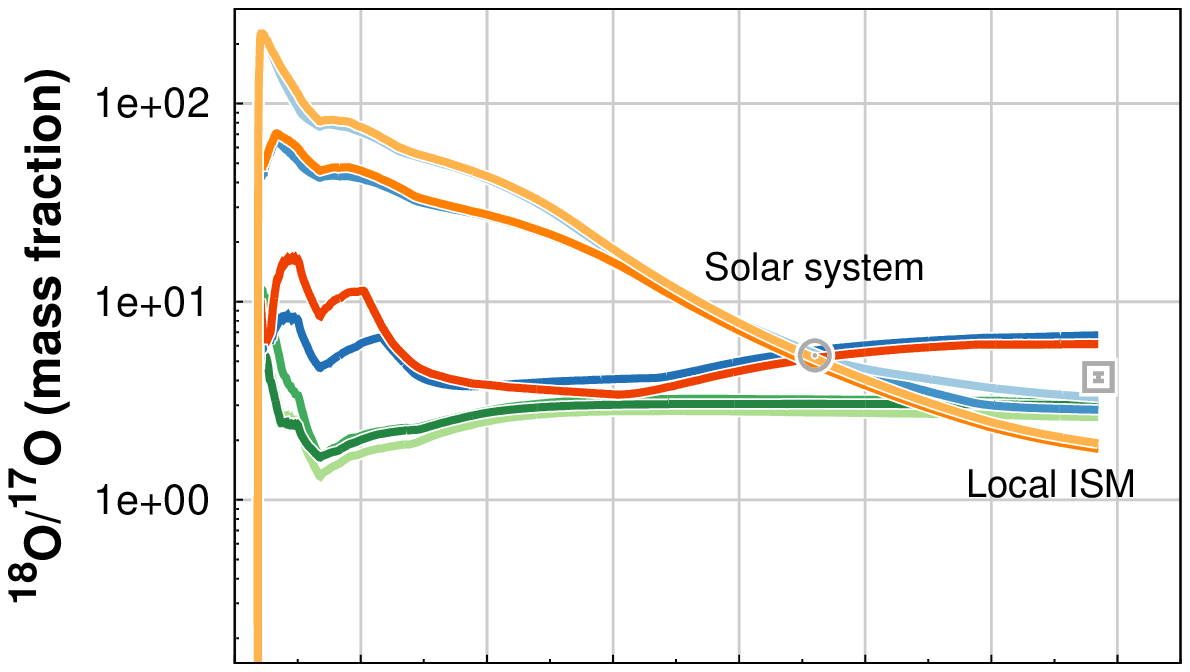}
\includegraphics[width=6.435cm]{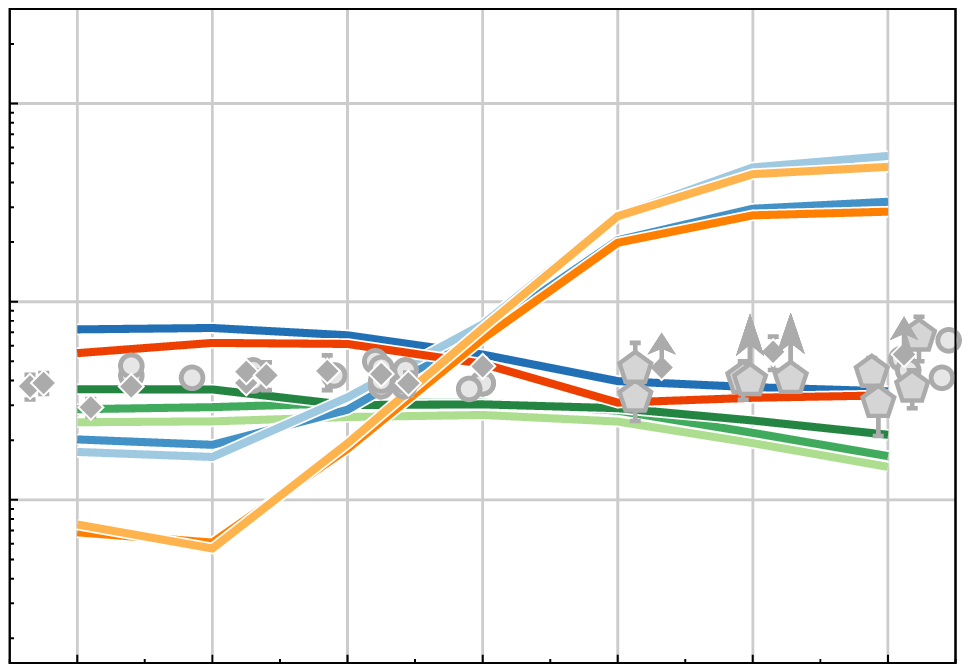}

\vspace{-0.35cm}

\includegraphics[width=7.8cm]{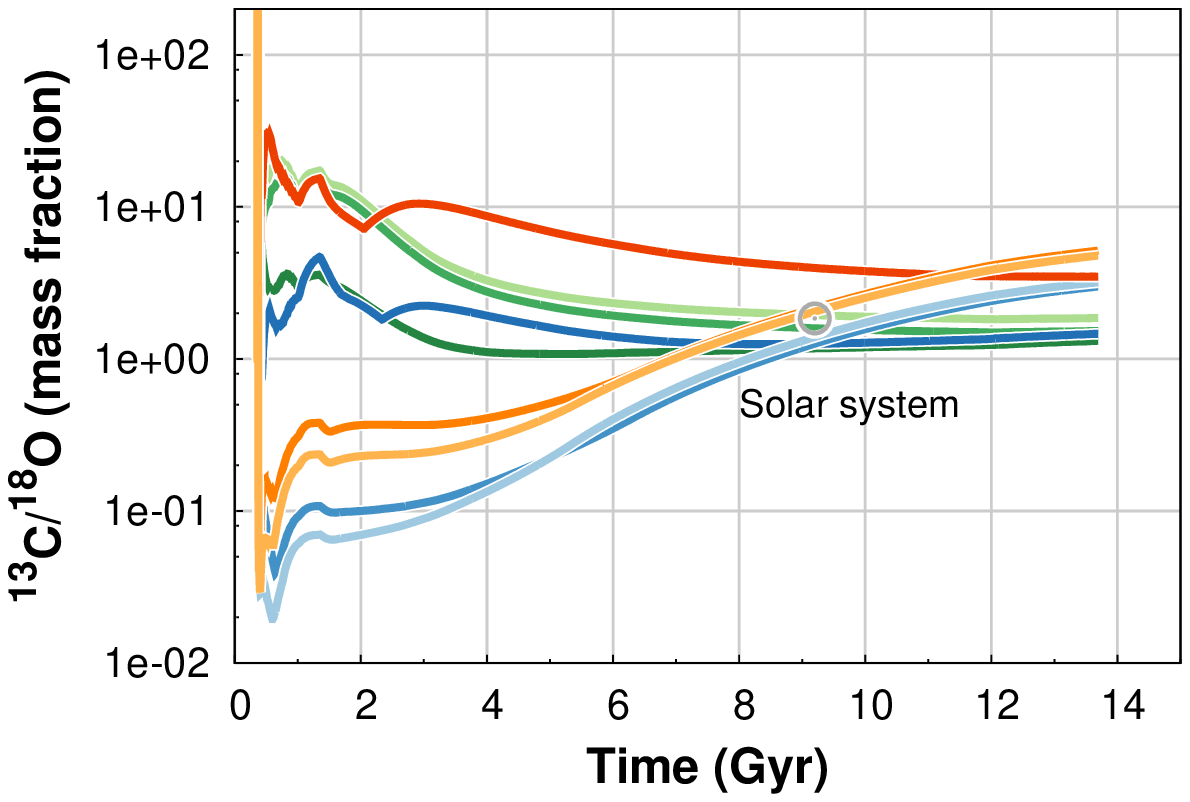}
\includegraphics[width=6.435cm]{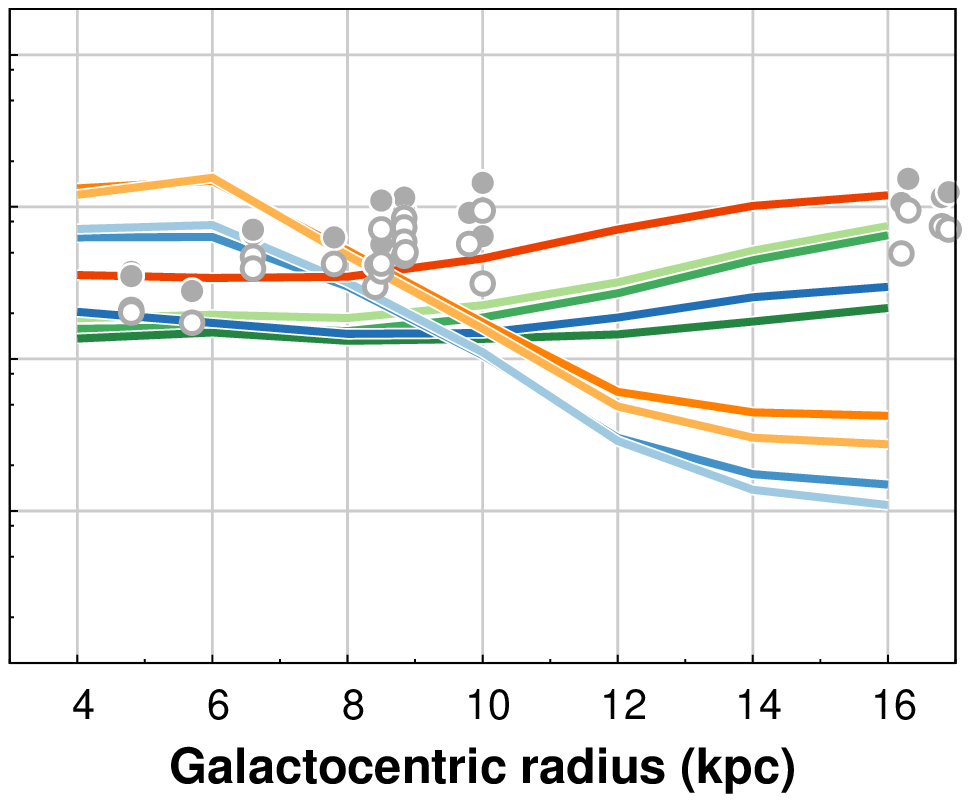} 

\caption{ \emph{Left-hand panels:} evolution of different CNO isotope ratios in 
  the solar vicinity. \emph{Right-hand panels:} corresponding present-time 
  gradients across the Milky Way disc. Theoretical predictions (solid lines) 
  are color-coded according to the adopted nucleosynthesis prescriptions (see 
  Table~\ref{tab:nuc} and explanatory legend in Fig.~\ref{fig:CNO}). Data (grey 
  symbols) sources are given in the text.}
\label{fig:MW}
\end{center}
\end{figure*}


With all the above-mentioned possible shortcomings in mind, we compare the 
trends predicted by our GCE model under different nucleosynthesis prescriptions 
(colored lines in Fig.~\ref{fig:CNO}) to the observational data from 
\citet{2004A&A...421..649I}, \citet{2005A&A...430..655S}, 
\citet{2014A&A...568A..25N}, \citet{2014AJ....147..136R}, 
\citet{2016A&A...591A..69S}\footnote{The O abundances for the stars in the 
  sample of \citet{2016A&A...591A..69S} are taken from 
\citet{2015A&A...576A..89B}.} and \citet{2019A&A...622L...4A} discussed in the 
previous paragraphs. Overall, the most appropriate yields appear to be those of 
\cite{2018ApJS..237...13L} for rotating massive stars coupled to those of 
\citet{2013MNRAS.431.3642V} for LIMS and super-AGB stars, i.e. models MWG-05 
and MWG-06. In particular, the relatively low N yields by 
\citet{2013MNRAS.431.3642V} guarantee a better fit to the observed [N/O] ratios 
for [O/H]~$> -$1.2, that is when the contributions from super-AGB stars first, 
and LIMS right after, start to become relevant.

\subsubsection{CNO isotope ratios in space and time in the Galaxy}

The determination of isotopic abundances in stars is limited to a handful of 
chemical species. For CNO elements, it can be made mostly in brilliant, giant 
stars. These, however, can have the original composition of their atmospheres 
altered by internal processes. It is thus convenient to resort to measurements 
of gas-phase isotopic abundances in interstellar clouds, though they provide 
only snapshots in time and, hence, no information about the past evolution that 
led to the observed configuration.

Radio observations are the prime tools for exploring the present-day gradients 
of $^{12}$C/$^{13}$C, $^{14}$N/$^{15}$N, $^{16}$O/$^{18}$O, $^{18}$O/$^{17}$O and 
$^{13}$C/$^{18}$O in the disc of our Galaxy. For the purpose of comparison with 
the predictions of our GCE model, we take the data from: (i) the compilation of 
\citet{1994ARA&A..32..191W}, based on results from molecular transitions of 
carbon monoxide and formaldehyde (small filled triangles in Fig.~\ref{fig:MW}, 
upper and middle right-hand panels); (ii) \citet{2005ApJ...634.1126M}, based on 
$N =$~1$-$0 transitions of the CN radical (large empty triangles in 
Fig.~\ref{fig:MW}, upper right-hand panel); (iii) \citet{2005A&A...437..957P}, 
which analysed the ground state rotational lines of the OH isotopologues (large 
empty circles in Fig.~\ref{fig:MW}, middle right-hand panel); (iv) 
\citet{2008A&A...487..237W}, providing a) the $^{13}$CO/C$^{18}$O gradient, from 
simultaneously observed $^{13}$CO and C$^{18}$O $J =$~1$-$0 and $J =$~2$-$1 
transitions (respectively, small filled and small empty circles in 
Fig.~\ref{fig:MW}, middle and lower right-hand panels; the original data have 
been combined with the $^{12}$C/$^{13}$C gradient by 
\citealt{2005ApJ...634.1126M} to trace the $^{16}$O/$^{18}$O gradient) and b) 
the $^{18}$O/$^{17}$O gradient, from $J =$~1$-$0, 2$-$1 and 3$-$2 transitions of 
C$^{18}$O and C$^{17}$O (small light-grey circles in Fig.~\ref{fig:MW}); (v) 
\citet{2016RAA....16...47L}, which derived C$^{18}$O/C$^{17}$O abundance ratios 
for 13 sources spanning the Galactocentric distance range 3--16~kpc (small 
filled diamonds in Fig.~\ref{fig:MW}); and (vi) \citet{2018MNRAS.478.3693C}, 
which targeted the $J =$~1$-$0 rotational transitions of hydrogen isocyanide 
and hydrogen cyanide and derived the $^{14}$N/$^{15}$N ratios for both molecules 
(Fig.~\ref{fig:MW}, the small filled and small empty pentagons are for HNC and 
HCN, respectively). To these we add some solid-state $^{12}$C/$^{13}$C data by 
\citet[][from interstellar CO$_2$ ices -- small empty squares in 
Fig.~\ref{fig:MW}, upper right-hand panel]{2000A&A...353..349B} and preliminary 
$^{12}$C/$^{13}$C and $^{18}$O/$^{17}$O measurements and lower limits for outer 
disc targets from Zhang et al. (2019, in preparation, large grey pentagons in 
Fig.~\ref{fig:MW}).

While the shapes of the present-day gradients of CNO isotope ratios in our 
Galaxy are reasonably well-known (with the exception of the outer disc, where 
only a few measurements are available; but see Zhang et al. 2019, in 
preparation), the knowledge of the evolution of these ratios in the solar 
neighbourhood is tightly limited. For the comparison with our model 
predictions, we rely on solar ratios from \citet[][$^{12}$C/$^{13}$C~$= 91.4 \pm 
1.3$, $^{16}$O/$^{18}$O~$= 511 \pm 10$, $^{18}$O/$^{17}$O~$= 5.36 \pm 
0.34$]{2013ApJ...765...46A} and \citet[][$^{14}$N/$^{15}$N~$= 441 \pm 
6$]{2011Sci...332.1533M}. These are shown as Sun symbols in Fig.~\ref{fig:MW}, 
left-hand panels. It should be noted, however, that the Sun might have migrated 
to its actual position from an inner birthplace \citep{1996A&A...314..438W} 
and, thus, its chemical composition might not reflect that of the local medium 
4.5~Gyr ago. For the local interstellar medium, we adopt average values (large 
open squares in Fig.~\ref{fig:MW}, left-hand panels) from the studies discussed 
above, namely: $\langle ^{12}$C/$^{13}$C$\rangle_{\rm{LISM}} = 68 \pm 15$ 
\citep{2005ApJ...634.1126M}, 
$\langle ^{16}$O/$^{18}$O$\rangle_{\rm{LISM}} = 395 \pm 56$ 
\citep{2005A&A...437..957P}, 
$\langle^{18}$O/$^{17}$O$\rangle_{\rm{LISM}} = 4.16 \pm 0.09$ 
\citep{2008A&A...487..237W}. For $\langle^{14}$N/$^{15}$N$\rangle_{\rm{LISM}}$, 
both the low value from \citet[][$290 \pm 40$]{2012ApJ...744..194A} and the 
high value from \citet[][$\sim 400$]{2018MNRAS.478.3693C} are displayed.

The predictions from models~MWG-01, MWG-02 and MWG-03 are in good agreement 
with the available $^{12}$C/$^{13}$C data both for the solar vicinity and across 
the disc. Should a substantial contribution to $^{13}$C production come from 
nova systems (see next section), models~MWG-05, MWG-06, MWG-07, MWG-08, MWG-09 
and MWG-10 could be made consistent with the observations, but the fit would 
worsen for models from MWG-01 to MWG-03.

Actually, in order to explain the observed decrease of the $^{14}$N/$^{15}$N 
ratio in the local disc over the last 4.5~Gyr, as well as its positive gradient 
across the disc, it is necessary that a large amount of $^{15}$N is supplied by 
some up to now neglected stellar factory. Thermonuclear runaways occurring in 
nova outbursts stand out as possible candidates \citep[but 
see][]{1999ApJ...512L.143C,2015ApJ...808L..43P}.

As regards the O isotope ratios, models in which the yields from fast-rotating 
massive stars are implemented during the full Galactic evolution -- namely, 
models~MWG-05, MWG-06, MWG-08 and MWG-09 -- produce gradients at variance with 
the observations. Indeed, some theoretical arguments 
\citep[see][]{1997A&A...321..465M} suggest that low-metallicity stars rotate 
faster than high-metallicity stars, on an average. For this reason, 
\citet{2018MNRAS.476.3432P} introduced an initial distribution of stellar 
rotational velocities (IDROV) in their GCE model, in analogy with the adoption 
of a stellar IMF. Because of the extremely loose constraints that can be 
imposed on the actual IDROV, however, their choice is far from unique. In the 
following section, we discuss the results of our GCE model when a simple step 
function is adopted for the IDROV: we assume the yields corresponding to 
stellar models computed with $\vel_{\rm{rot}} = 300$~km s$^{-1}$ for [Fe/H]~$< -$1 
and the yields of non-rotating models at higher metallicities. In the latter 
case, we also set to 60--100~M$_\odot$, rather than 35--100~M$_\odot$, the mass 
range for full collapse to black holes, which guarantees a better fit to the 
observed O abundances of local disc stars in the framework of our model.

\subsubsection{A possible, comprehensive evolutive scenario}
\label{sec:scen}


\begin{figure*}
\begin{center}
\hspace{1.1cm}
\includegraphics[width=15.92cm]{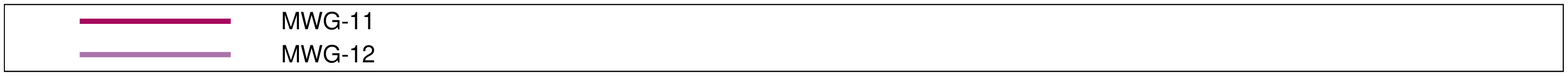}

\includegraphics[width=7.8cm]{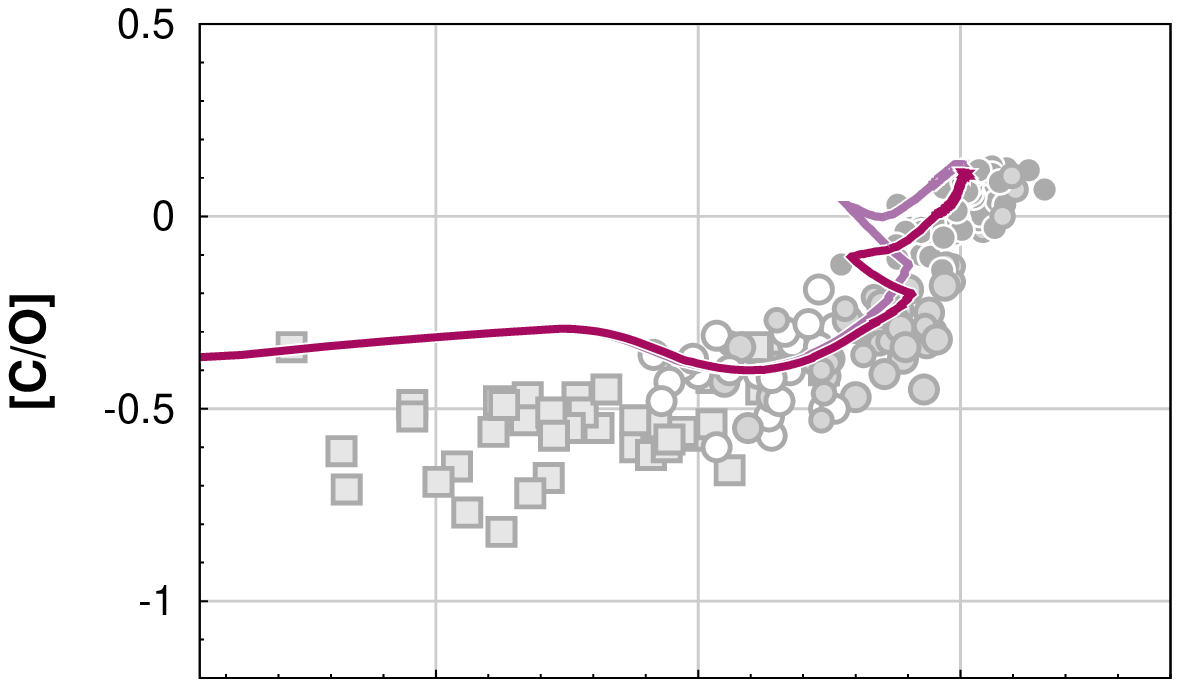}
\hspace{1cm}
\includegraphics[width=7.8cm]{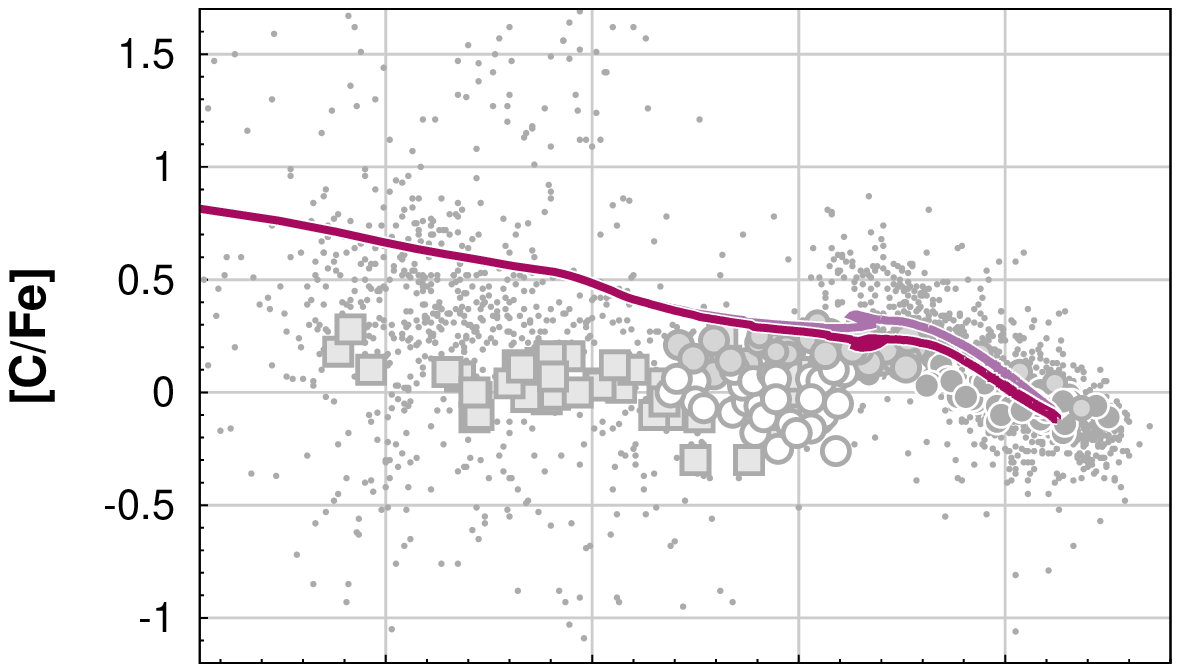}

\vspace{-0.35cm}

\includegraphics[width=7.8cm]{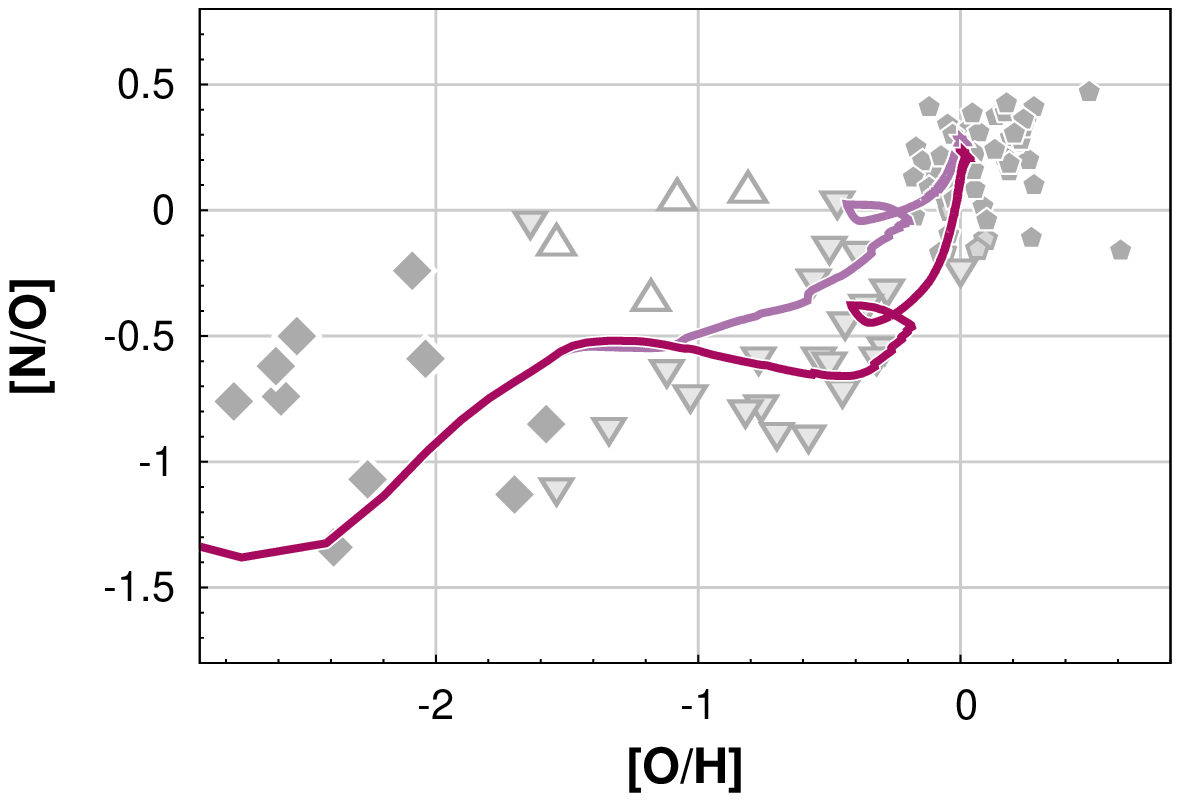}
\hspace{1cm}
\includegraphics[width=7.8cm]{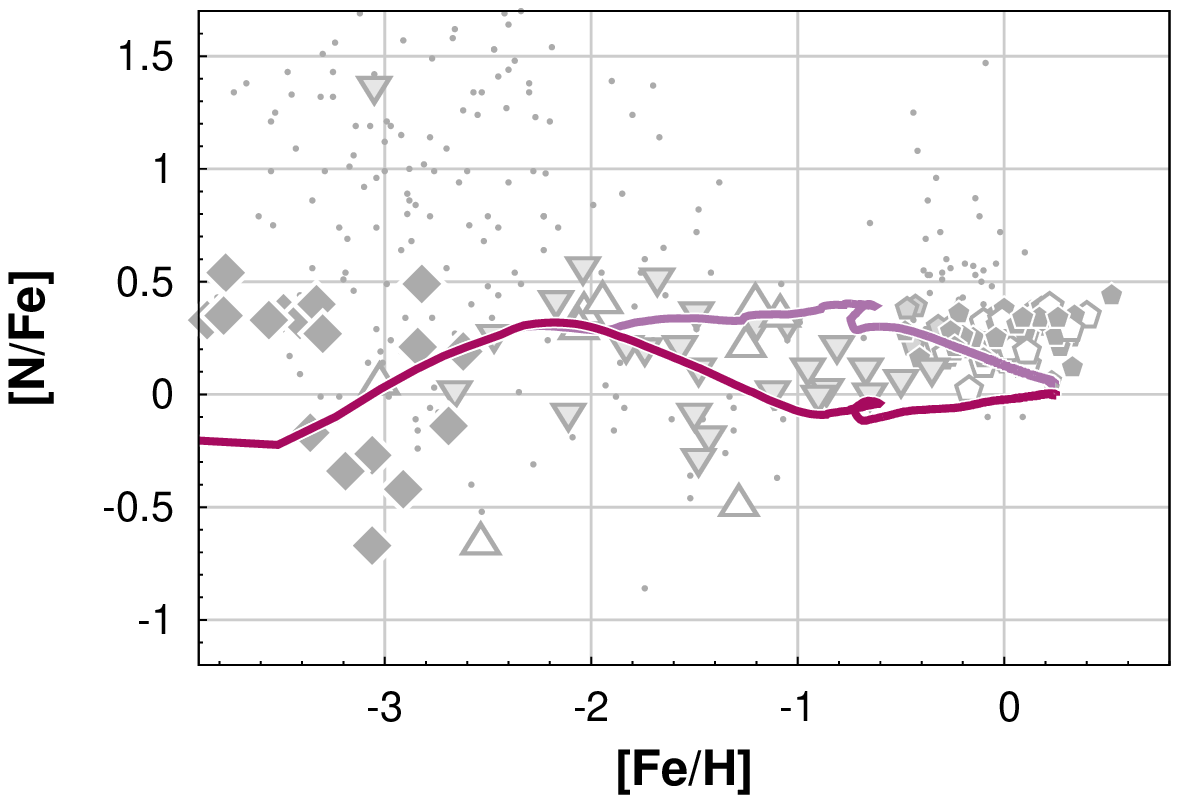}

\caption{ Same as Fig.~\ref{fig:CNO}, but the model predictions are from models 
  MWG-11 and MWG-12 (bordeaux and lilac lines, respectively).}
\label{fig:CNOnov}
\end{center}
\end{figure*}



\begin{figure}
\begin{center}
\includegraphics[width=7.8cm]{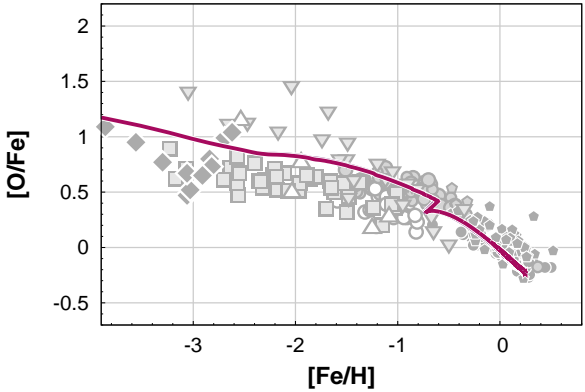}
\caption{ Same as Fig.~\ref{fig:O}, but shown are predictions from models 
  MWG-11 and MWG-12 (the latter, concealed behind model~MWG-11 predictions).}
\label{fig:Onov}
\end{center}
\end{figure}



\begin{figure*}
\begin{center}
\includegraphics[width=7.8cm]{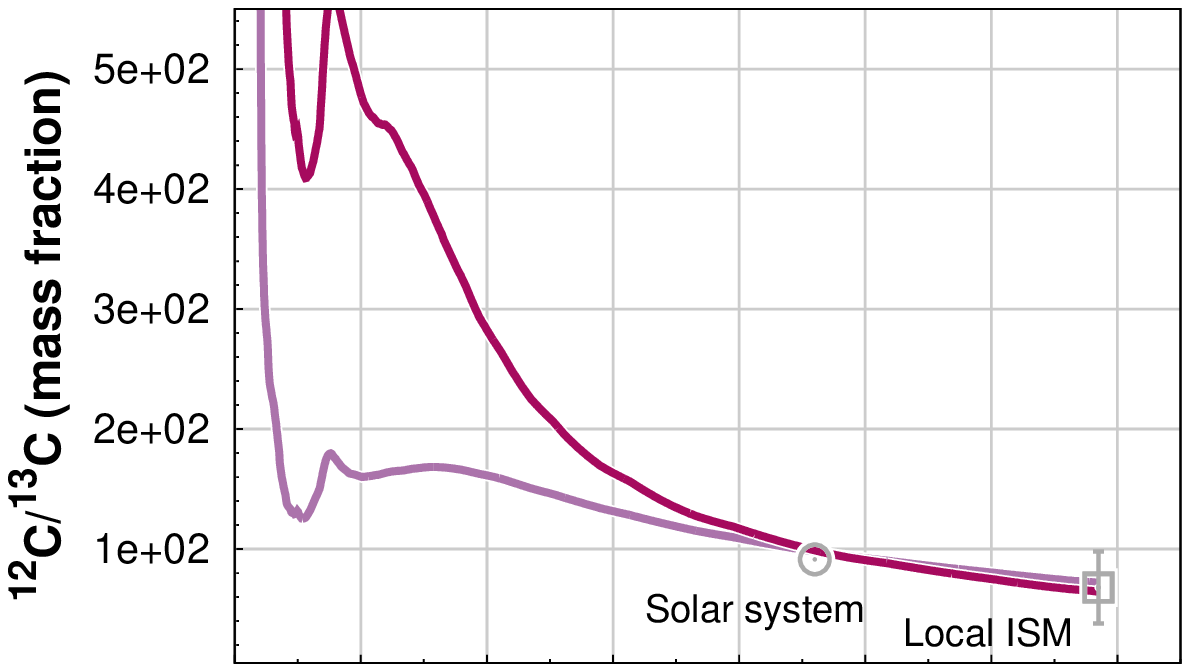}
\includegraphics[width=6.435cm]{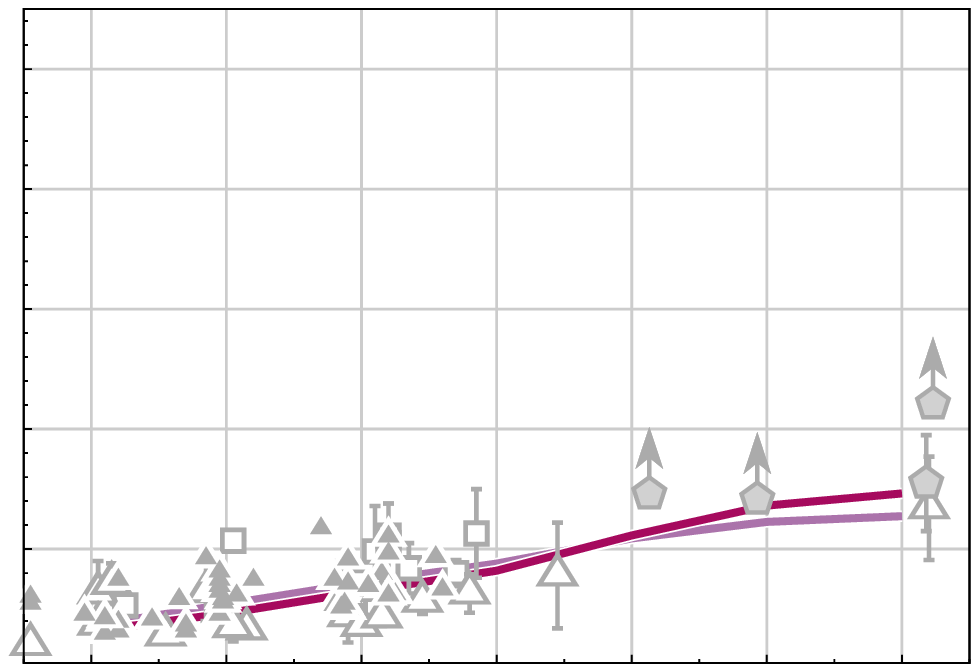}

\vspace{-0.35cm}

\includegraphics[width=7.8cm]{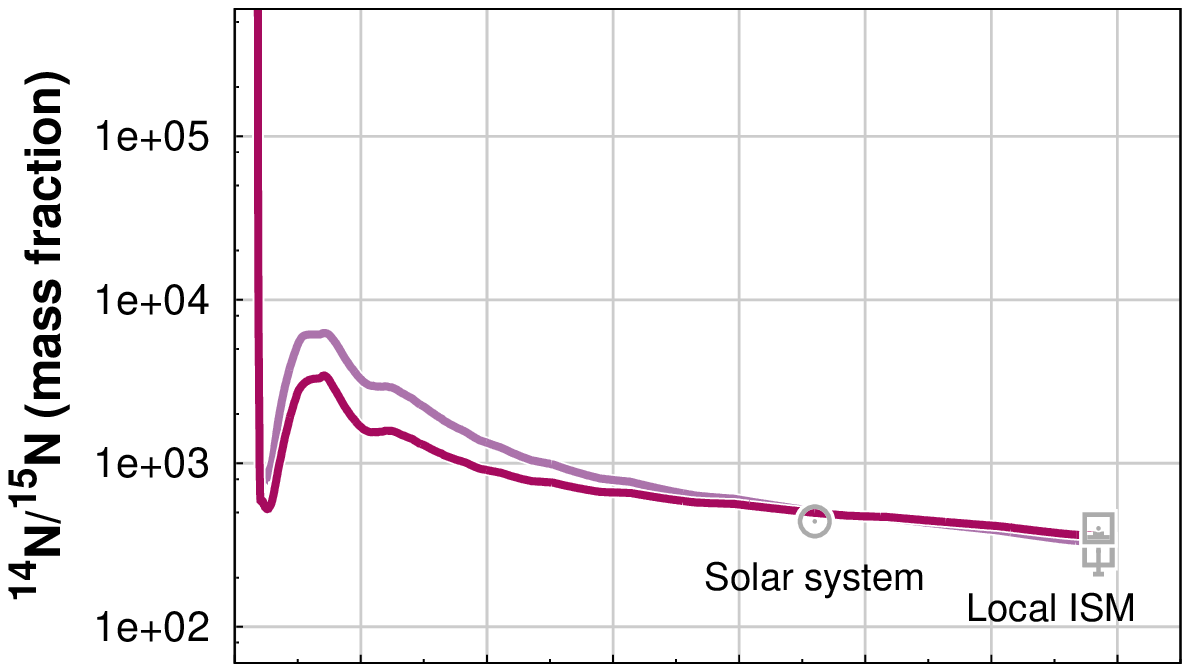}
\includegraphics[width=6.435cm]{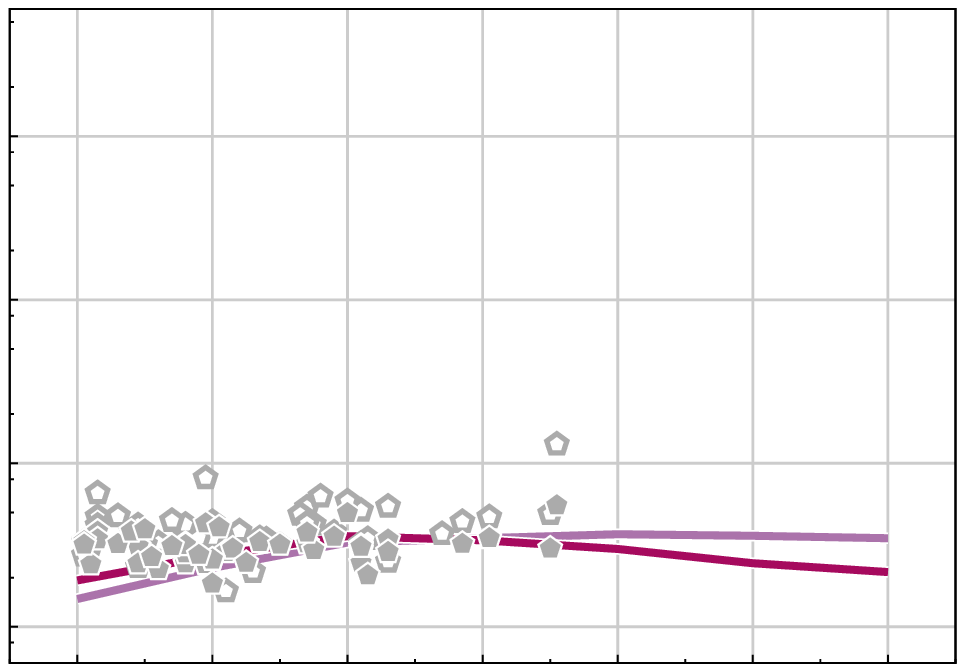}

\vspace{-0.35cm}

\includegraphics[width=7.8cm]{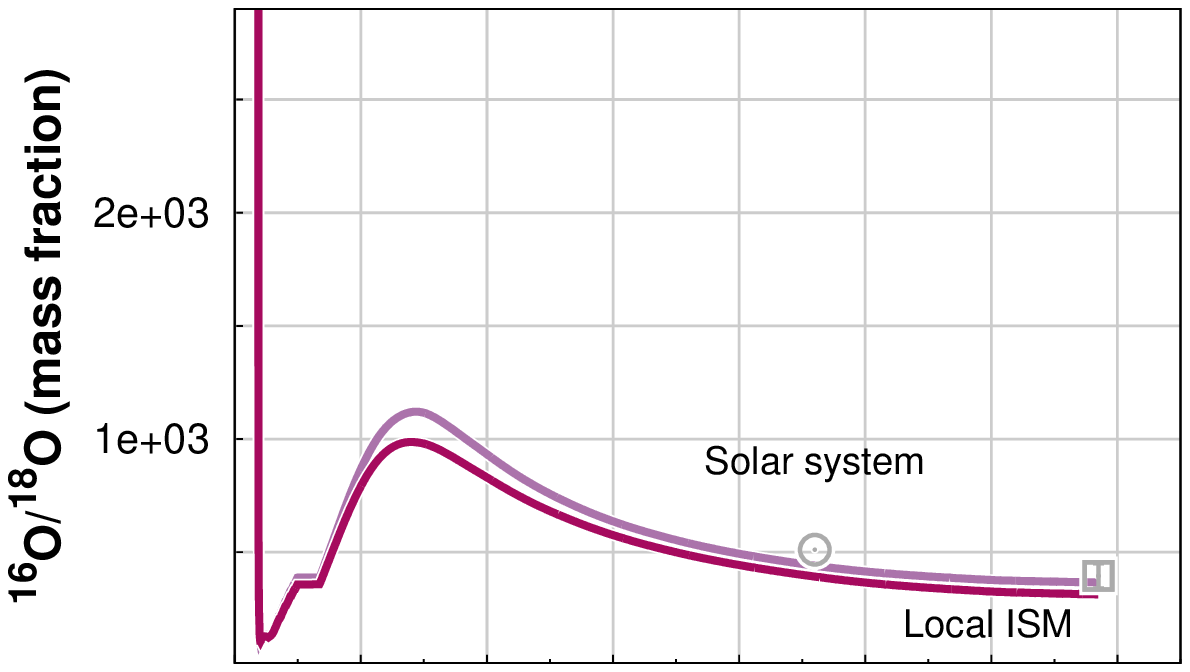}
\includegraphics[width=6.435cm]{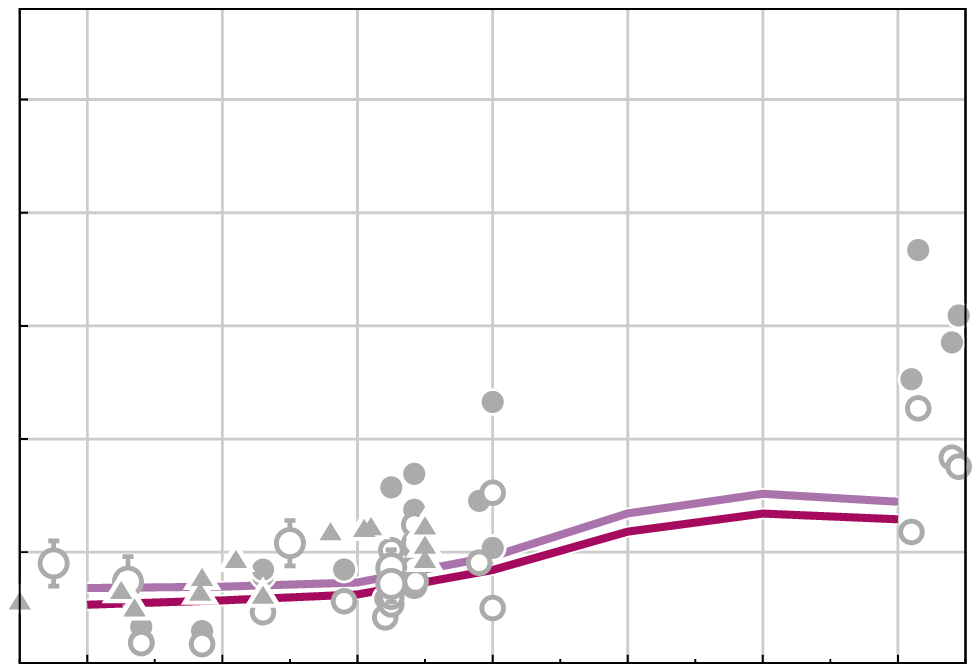}

\vspace{-0.35cm}

\includegraphics[width=7.8cm]{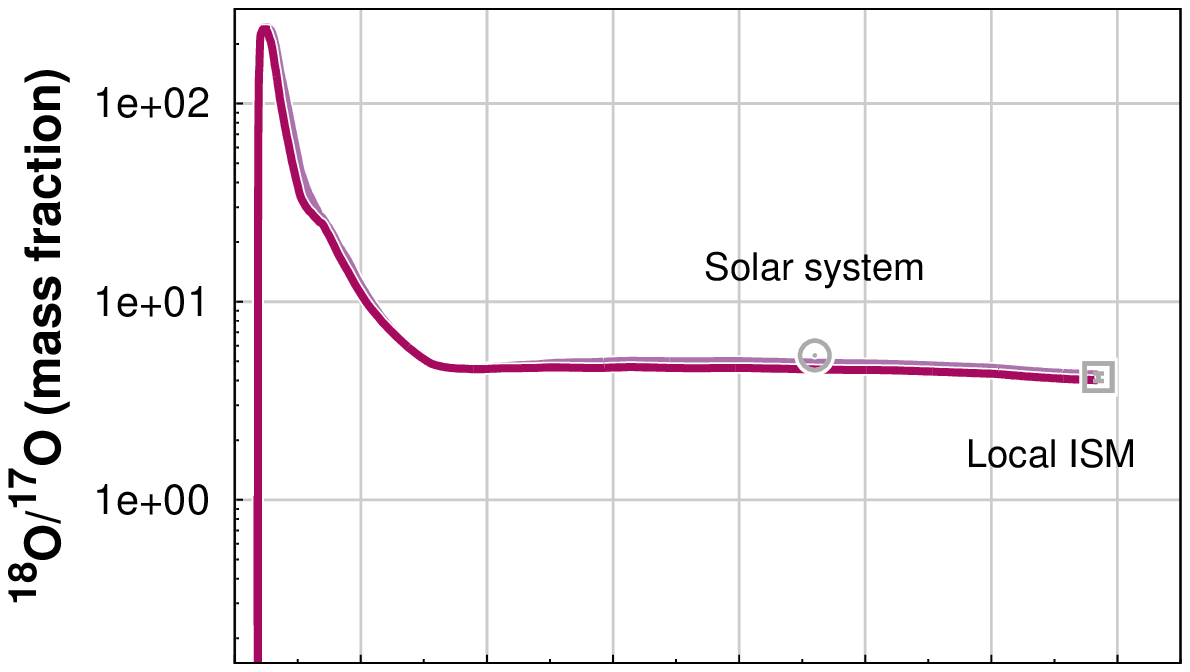}
\includegraphics[width=6.435cm]{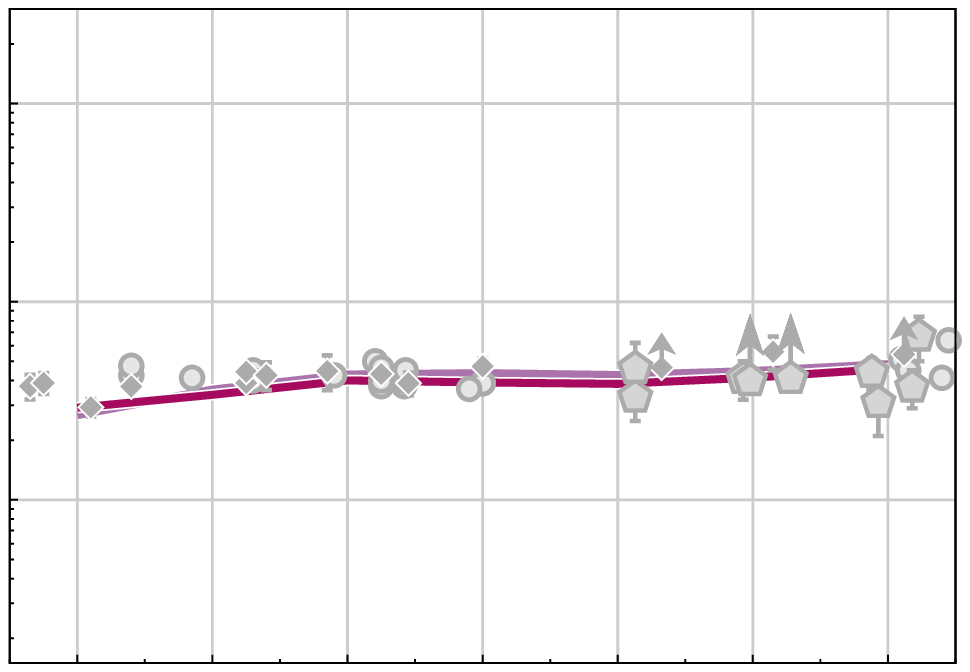}

\vspace{-0.35cm}

\includegraphics[width=7.8cm]{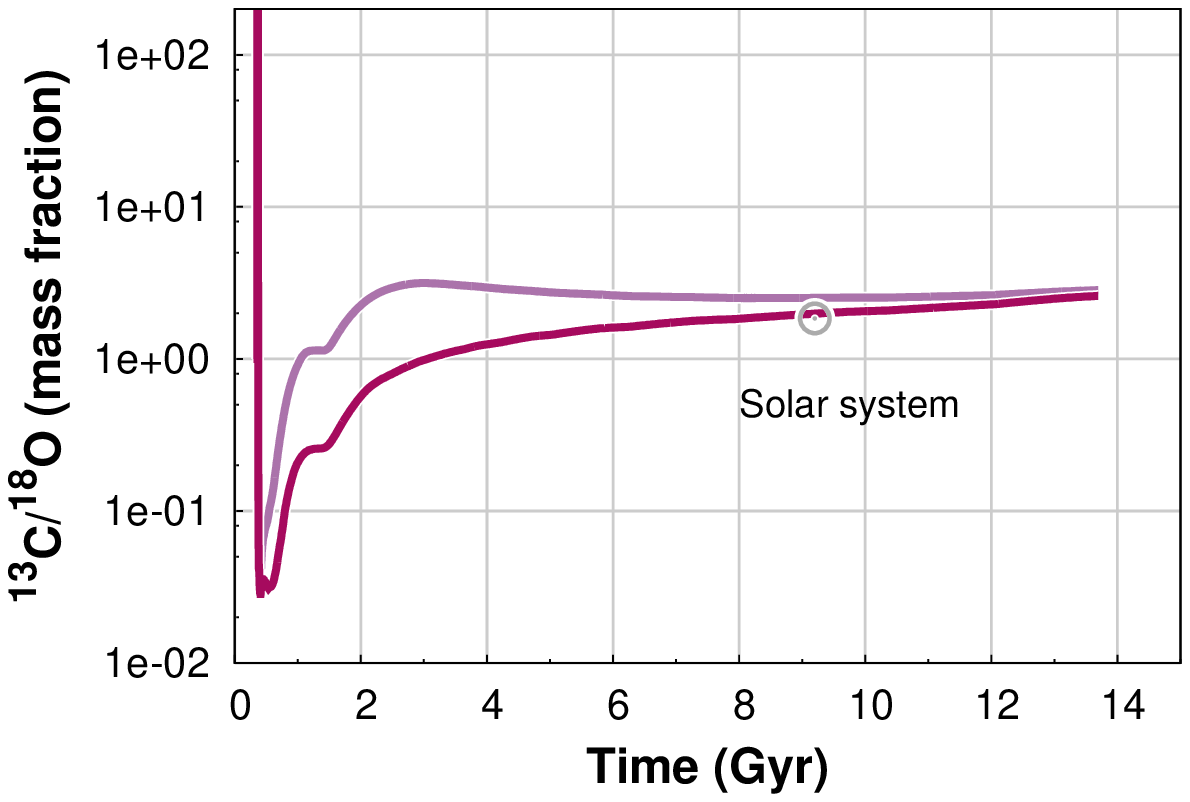}
\includegraphics[width=6.435cm]{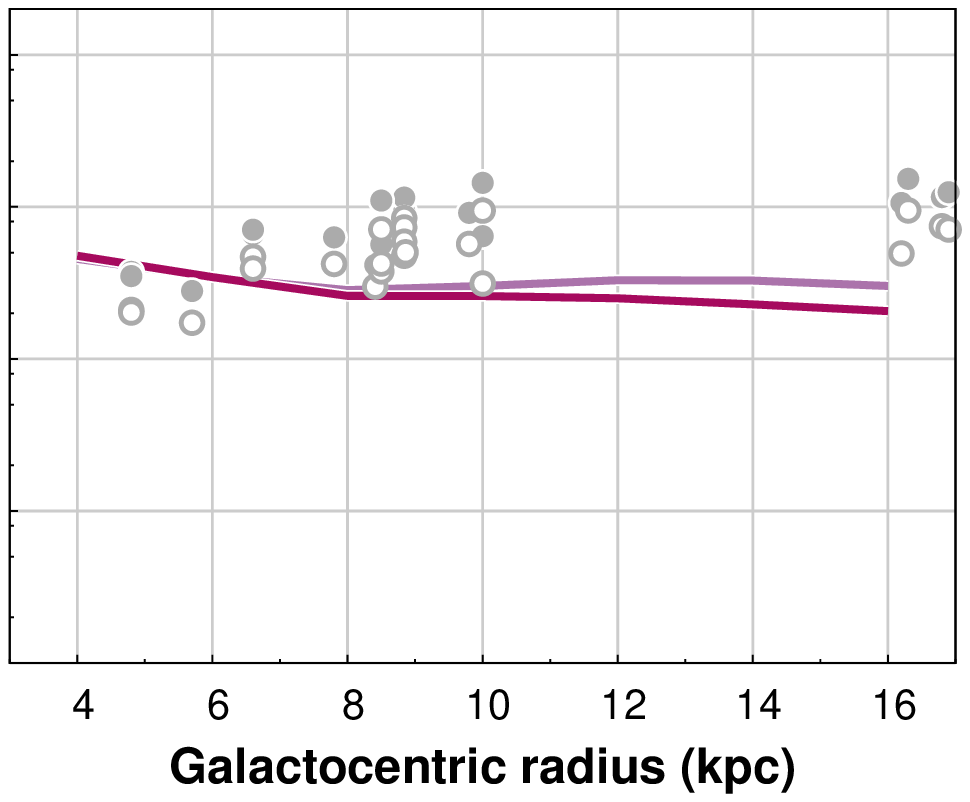} 

\caption{ Same as Fig.~\ref{fig:MW}, but shown are predictions from models 
  MWG-11 and MWG-12 (bordeaux and lilac lines, respectively).}
\label{fig:MWnov}
\end{center}
\end{figure*}


From Figs.~\ref{fig:CNO} and \ref{fig:O} discussed above, one sees that the 
adoption of massive star yields computed with either $\vel_{\rm{rot}} = 300$~km 
s$^{-1}$ or $\vel_{\rm{rot}} = 150$~km s$^{-1}$ results in comparable [C/Fe], [C/O] 
and [O/Fe] ratios, that differ by 0.15~dex at most, during the entire solar 
neighbourhood evolution. Larger differences (up to 0.5~dex at the lowest 
metallicities) are found between the theoretical [N/Fe] and [N/O] ratios 
predicted for halo stars in the two cases, while the differences cancel out at 
disc metallicities.

The largest contrasts arise anyway between the evolutive tracks obtained in the 
rotating case (whatever the value of $\vel_{\rm{rot}}$) and the non-rotating one. 
While nothing can be said from direct observations about the distribution of 
the rotational velocities of low-metallicity, high-mass stars, there is strong, 
indirect evidence from the chemical composition of unevolved halo stars that 
below [Fe/H]~$= -$2 most SNII progenitors must have been fast rotators: indeed, 
fast rotation leads to the synthesis of large amounts of primary $^{14}$N at 
low metallicities, which explains the relevant observations \citep[see 
also][and references therein]{2006A&A...449L..27C,2018MNRAS.476.3432P}.

Let us now throw the minor CNO isotopes into the mix (Fig.~\ref{fig:MW}): the 
situation gets more complicated! In fact, novae -- not included in models from 
MWG-01 to MWG-10 -- may inject significant amounts of $^{13}$C, $^{15}$N and 
$^{17}$O in the Galactic medium on long time scales \citep[see][and references 
therein]{2003MNRAS.342..185R}. If implemented in models including fast 
rotators, this contribution may, in principle, reverse the trend at odds with 
the observations predicted for $^{14}$N/$^{15}$N and improve the agreement 
between predicted and observed $^{12}$C/$^{13}$C ratios. Yet, the 
$^{16}$O/$^{18}$O ratio is immune to changes caused by an implementation of this 
nucleosynthetic channel (neither $^{16}$O nor $^{18}$O are produced in 
appreciable amounts in nova outbursts).

From Fig.~\ref{fig:MW}, middle panels, it can be seen that GCE models without 
fast rotators produce $^{16}$O/$^{18}$O ratios in remarkable agreement with 
local and inner disc data. From the more uncertain abundance determinations in 
the outer disc, however, it can not be excluded that a fraction of massive 
stars are rotating fast at large Galactocentric distances. Our GCE model 
predicts the current metallicity in the inner (outer) Milky Way disc to be 
$Z \simeq$~0.016--0.028 (0.005--0.009) at $R_{\mathrm{G}} =$ 4 (16)~kpc (the 
exact value depending on the choice of the stellar yields). Thus, the overall 
picture is one in which the formation of fast-rotating massive stars is 
inhibited in environments with solar or super-solar metallicity, while it can 
not be excluded that a fraction of massive stars are rotating fast in regions 
with metallicities close to those typical of the Magellanic Clouds. Likewise 
from N observations, and as expected from theory \citep{1997A&A...321..465M}, 
from the $^{16}$O/$^{18}$O ratios we get an indication that fast-rotating 
massive stars are more common in less evolved environments. It goes without 
saying that more precise measurements of the $^{16}$O/$^{18}$O ratios in more 
populous samples of low-metallicity molecular clouds are crucial to better 
understand the transition from the fast-rotation to the null(or slow)-rotation 
regime. It is worth recalling at this point that the $^{16}$O/$^{18}$O gradient 
in the Milky Way is determined from $^{13}$CO and C$^{18}$O transitions 
\citep{2008A&A...487..237W} via the adoption of a $^{12}$C/$^{13}$C gradient 
that is poorly defined in the outer disc (see Fig.~\ref{fig:CNO}, upper 
right-hand panel). Thus, better $^{16}$O/$^{18}$O determinations call for better 
estimates of the $^{12}$C/$^{13}$C ratio. We will address this issue in detail 
in a forthcoming paper (Zhang et al. 2019, in preparation). Grids of yields 
computed for rotational velocities in between those considered here are also in 
demand.


\begin{figure*}
\begin{center}
\includegraphics[width=\textwidth]{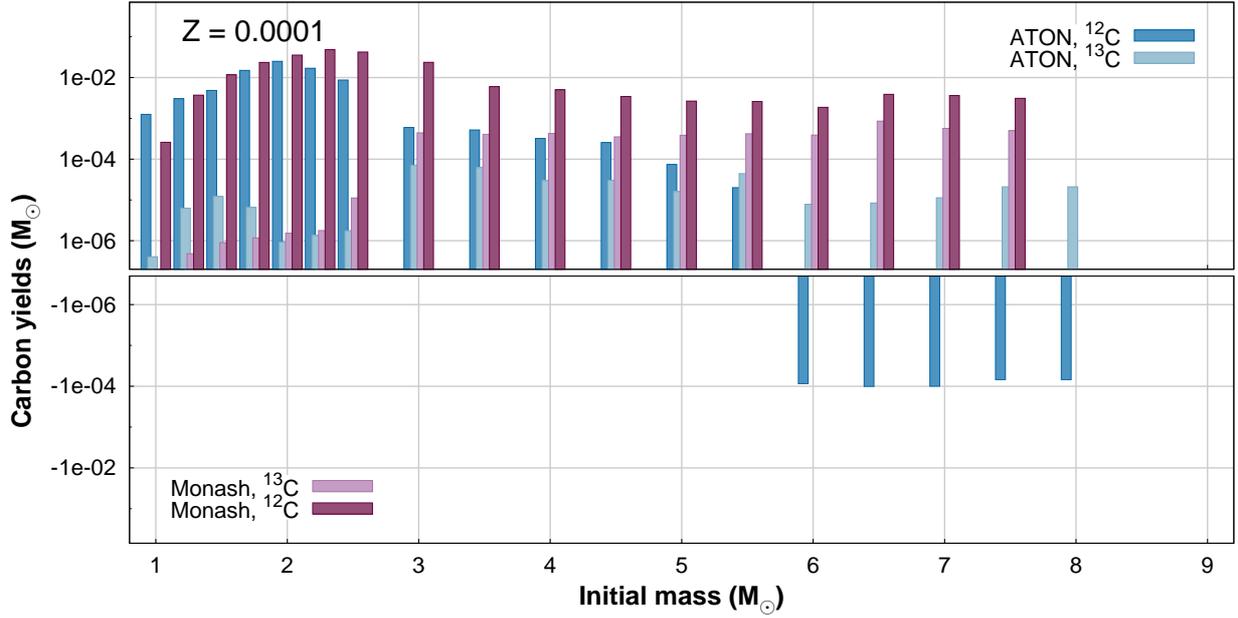}
\caption{ Net yields of $^{12}$C and $^{13}$C from LIMS and super-AGB stars 
  after \citet{2013MNRAS.431.3642V}, in shades of blue, and 
  \citet{2010MNRAS.403.1413K} and 
  \citet{2014MNRAS.437..195D,2014MNRAS.441..582D}, in shades of purple, for 
  $Z = 0.0001$. A negative yield means the element is destroyed, rather than 
  produced, in the star.}
\label{fig:CylowZ}
\end{center}
\end{figure*}



\begin{figure*}
\begin{center}
\includegraphics[width=\textwidth]{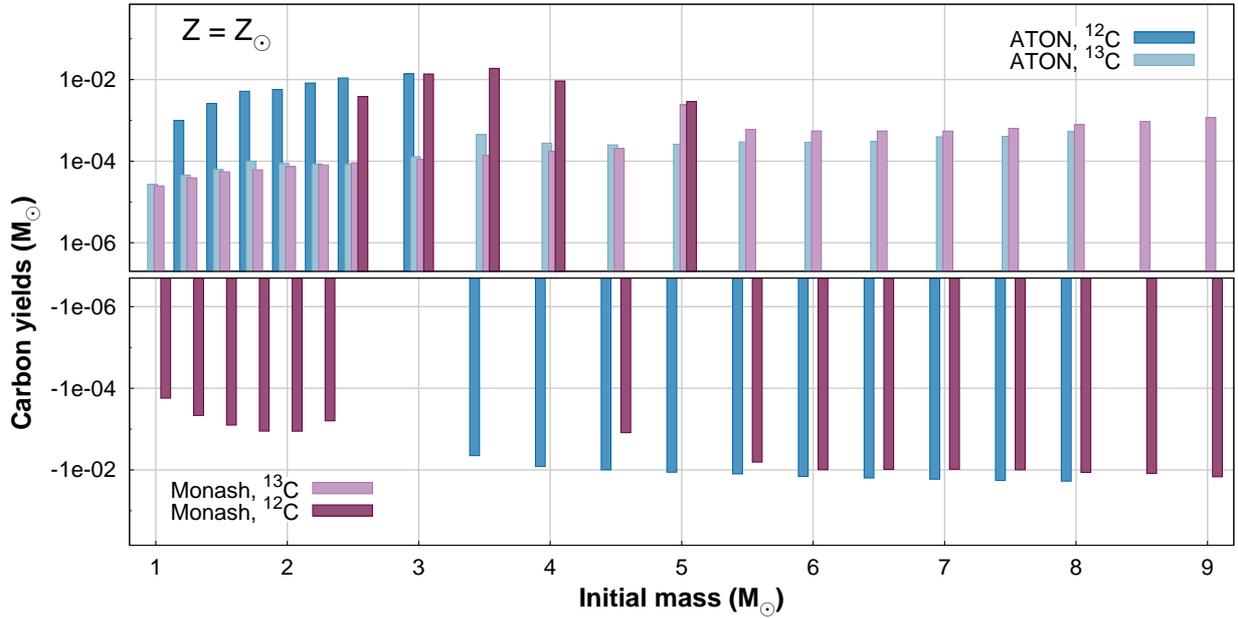}
\caption{ Same as Fig.~\ref{fig:CylowZ}, but for solar metallicity.}
\label{fig:CyZsun}
\end{center}
\end{figure*}


In the remainder of this section we use what is at hand and discuss the results 
of two GCE models that: (i) assume a simple step function for the IDROV of 
core-collapse SN progenitors and (ii) take into account the production of rare 
isotopes during nova outbursts.

As regards the massive stars, for [Fe/H]~$= -$3 and [Fe/H]~$= -$2 we adopt the 
recommended yields by \citet[][their set R]{2018ApJS..237...13L} computed with 
$\vel_{\rm{rot}} =$ 300~km s$^{-1}$. For [Fe/H]~$= -$1 and [Fe/H]~=~0, we use the 
yields for non-rotating stars and reduce the mass range for full collapse to 
black holes from 30--100~M$_\odot$ to 60--100~M$_\odot$, which improves the 
predictions about the behaviour of [O/Fe] versus [Fe/H] -- and, hence, the run 
of [C/O] and [N/O] as a function of [O/H] -- at disc metallicities in our 
models. The yields for LIMS and super-AGB stars are taken from \citet[][model 
MWG-11]{2013MNRAS.431.3642V} or from \citet{2010MNRAS.403.1413K} and 
\citet[][model MWG-12]{2014MNRAS.437..195D,2014MNRAS.441..582D}. Linear 
interpolation in mass and metallicity is performed between adjacent values of 
the yields. In both model~MWG-11 and model~MWG-12, nova nucleosynthesis is 
implemented following the prescriptions of \citet[][and references 
therein; see Section~\ref{sec:bs}]{1999A&A...352..117R,2017MNRAS.470..401R}.

The results of models~MWG-11 and MWG-12 are illustrated in 
Figs.~\ref{fig:CNOnov}--\ref{fig:MWnov}. From Figs.~\ref{fig:CNOnov} and 
\ref{fig:Onov}, it can be seen that for the main CNO isotopes the agreement 
with the observations is from satisfactory to very good during most of the 
solar neighbourhood evolution. The predicted overabundances of carbon and (to a 
lower extent) oxygen in the early Galaxy have to be expected and could be 
rectified by the inclusion of a fraction of hypernovae at low metallicities 
(see the discussion in Section~\ref{sec:aburat}). Such an energy effect would 
also have the advantage of redressing the well-known discrepancies between 
predicted and observed cobalt and zinc abundances in low-metallicity Galactic 
halo stars \citep{2006ApJ...653.1145K,2010A&A...522A..32R,2018MNRAS.476.3432P}.

Stars with initial masses $m \le$~9~M$_\odot$ do not produce $^{16}$O (rather, 
they partly destroy it in their interiors), but synthesise about one third of 
the solar $^{12}$C (30 per cent for model MWG-11, and more than 35 per cent for 
model MWG-12), as well as most of the solar $^{14}$N (65 per cent in the case 
of model MWG-11, and more than 70 per cent for model MWG-12). As already 
stressed by \citet{2005ApJ...623..213C}, the fractional contribution from LIMS 
(and super-AGB stars) and massive stars is strongly dependent on time, on the 
galactic region under scrutiny and, of course, on the adopted GCE model and 
stellar yields. Therefore, one should beware of any daring extrapolations of 
the rates reported above to other galaxies. In the next section, we further 
stress this point by discussing the results for a prototype SMG observed at 
high redshift.

The enrichment of $^{13}$C coarsely tracks that of $^{14}$N, with more than 50 
and 70 per cent of the solar $^{13}$C being produced by LIMS and super-AGB 
stars in models MWG-11 and MWG-12, respectively. In particular, according to 
\citet{2010MNRAS.403.1413K} and 
\citet{2014MNRAS.437..195D,2014MNRAS.441..582D}, low-metallicity intermediate- 
to high-mass AGB stars are effective $^{13}$C producers (see 
Fig.~\ref{fig:CylowZ}), which makes the predicted carbon isotope ratio to reach 
down to $^{12}$C/$^{13}$C~$\simeq$ 130 at the end of the halo/thick-disc 
formation in model MWG-12 (Fig.~\ref{fig:MWnov}, upper left-hand panel). 
Model~MWG-11, that assumes the much lower $^{13}$C yields by 
\citet{2013MNRAS.431.3642V} for the same stars, barely reaches down to 
$^{12}$C/$^{13}$C~$\sim$ 400, making it harder to conciliate its predictions to 
observations of post first dredge-up halo stars \citep{2006A&A...455..291S}. At 
solar metallicities, the $^{13}$C yields from both sets of stellar models align 
with each other and the stars with $m >$ 4~M$_\odot$ efficiently destroy $^{12}$C 
in their interiors (see Fig.~\ref{fig:CyZsun}), which leads to converging GCE 
results (the slightly more pronounced decline characterising the late evolution 
of the carbon isotope ratio in model~MWG-11 compared to model~MWG-12 is due to 
the slightly larger $^{13}$C nova yield necessary to bring the predicted solar 
abundance of $^{13}$C in agreement with the observed one in model~MWG-11).

The minor isotope of nitrogen, $^{15}$N, is destroyed in non-rotating stellar 
models, but it is efficiently synthesised in the presence of rotation, owing 
to the diffusion of matter between the H- and He-burning zones 
\citep{2018ApJS..237...13L,2018MNRAS.476.3432P}. However, this production is 
not sufficient, alone, to explain the Galactic $^{14}$N/$^{15}$N data. In this 
work, we fit well measurements of nitrogen isotope ratios in Milky Way targets 
by resorting to nova pollution (see Fig.~\ref{fig:MWnov}), but we ought to 
mention that unusually high $^{15}$N abundances are found in some Galactic AGB 
stars \citep{2013ApJ...768L..11H} and planetary nebulae 
\citep{2018Natur.564..378S}. These are currently unexplained by stellar 
evolution theory and might call for some substantial revision of stellar 
physics.

The $^{17}$O yields are overall positive, apart from those from $m >$ 
20--25~M$_\odot$ star models, that turn negative\footnote{This is true only for 
the yield sets by \cite{2018ApJS..237...13L}; \cite{2013ARA&A..51..457N} always 
report positive $^{17}$O yields for massive stars.} for [Fe/H]~$\ge -$1. As a 
consequence, the predicted $^{18}$O/$^{17}$O ratio decreases with time in the 
local disc, except in the last 4.5 Gyr, when it would mildly increase without 
the addition of a nova contribution. Our GCE models, including such 
contribution, account very well for the local evolution of this isotope ratio, 
as well as its Galactic gradient (see Fig.~\ref{fig:MWnov}, second left- and 
right-hand panels down, respectively).

As regards $^{18}$O, its yields are largely positive for stars in the mass 
interval 13--25~M$_\odot$, while out of this range the stars either destroy 
$^{18}$O, or provide smaller amounts of it. From Fig.~\ref{fig:MWnov}, middle 
panels, it is seen that the evolution of the $^{16}$O/$^{18}$O isotope ratio in 
the Milky Way is reproduced fairly well by both models~MWG-11 and MWG-12. 
However, in order to assess the validity of the models at relatively low 
metallicities, it is mandatory to obtain firmer measurements of this ratio in 
the external disc (see arguments set out in previous paragraphs).

\subsection{Prototype SMG}
\label{sec:smg}


\begin{table*}
\caption{Final$^a$ abundance ratios of CNO isotopes in the interstellar medium 
for our grid of SMG models.}
\begin{tabular}{@{}cccccccc@{}}
\hline
{Model} & SFH & IMF & $^{12}$C/$^{13}$C & $^{14}$N/$^{15}$N & $^{16}$O/$^{18}$O & $^{18}$O/$^{17}$O & $^{13}$C/$^{18}$O \\
\hline
\textcolor{lightgreen}{SMG-01} & Continuous, $\Delta t_{\mathrm{burst}} = 1$~Gyr & Canonical & 118 & $4.5 \times 10^4$ & $2.9 \times 10^3$ & 2.5 & 6.2 \\
                               & One-shot, $\Delta t_{\mathrm{burst}} = 20$~Myr & Canonical & $1.7 \times 10^3$ & $1.8 \times 10^6$ & $3.9 \times 10^4$ & 5.3 & 1.6 \\
                               & Gasping, $\Delta t_{\mathrm{burst}} = 50$~Myr & Canonical & 132 & $3.8 \times 10^4$ & $2.3 \times 10^3$ & 2.4 & 5 \\
                               & Gasping, $\Delta t_{\mathrm{burst}} = 50$~Myr & Top-heavy & 64 & 453 & 57 & 3.2 & 0.09 \\
 & & & & & & & \\
\textcolor{lightorange}{SMG-08} & Continuous, $\Delta t_{\mathrm{burst}} = 1$~Gyr & Canonical & 162 & 400 & 79 & 81 & 0.2 \\
                                & One-shot, $\Delta t_{\mathrm{burst}} = 20$~Myr & Canonical & 765 & $3.9 \times 10^3$ & 447 & 126 & 0.1 \\
                                & Gasping, $\Delta t_{\mathrm{burst}} = 50$~Myr & Canonical & 176 & 424 & 85 & 65 & 0.2 \\
                                & Gasping, $\Delta t_{\mathrm{burst}} = 50$~Myr & Top-heavy & 173 & $1.7 \times 10^4$ & 334 & 6 & 0.9 \\
 & & & & & & & \\
\textcolor{darkorange}{SMG-09} & Continuous, $\Delta t_{\mathrm{burst}} = 1$~Gyr & Canonical & 175 & 214 & 126 & 45 & 0.3 \\
                               & One-shot, $\Delta t_{\mathrm{burst}} = 20$~Myr & Canonical & 970 & 695 & 451 & 99 & 0.1 \\
                               & Gasping, $\Delta t_{\mathrm{burst}} = 50$~Myr & Canonical & 193 & 220 & 131 & 37 & 0.3 \\
                               & Gasping, $\Delta t_{\mathrm{burst}} = 50$~Myr & Top-heavy & 209 & $2.4 \times 10^3$ & 480 & 4 & 1.1 \\
 & & & & & & & \\
\textcolor{jazz}{SMG-11} & Continuous, $\Delta t_{\mathrm{burst}} = 1$~Gyr & Canonical & 521 & $2.7 \times 10^3$ & 344 & 1.4 & 0.23 \\
                         & One-shot, $\Delta t_{\mathrm{burst}} = 20$~Myr & Canonical & $1.3 \times 10^3$ & $1.3 \times 10^4$ & 819 & 168 & 0.13 \\
                         & Gasping, $\Delta t_{\mathrm{burst}} = 50$~Myr & Canonical & 546 & $3.0 \times 10^3$ & 372 & 1 & 0.3 \\
                         & Gasping, $\Delta t_{\mathrm{burst}} = 50$~Myr & Top-heavy & 307 & $6.3 \times 10^4$ & $2.8 \times 10^3$ & 0.01 & 3.6 \\
 & & & & & & & \\
\textcolor{lilac}{SMG-12} & Continuous, $\Delta t_{\mathrm{burst}} = 1$~Gyr & Canonical & 184 & $3.9 \times 10^3$ & 380 & 27 & 0.95 \\
                          & One-shot, $\Delta t_{\mathrm{burst}} = 20$~Myr & Canonical & $1.3 \times 10^3$ & $1.3 \times 10^4$ & 815 & 168 & 0.13 \\
                          & Gasping, $\Delta t_{\mathrm{burst}} = 50$~Myr & Canonical & 201 & $3.7 \times 10^3$ & 419 & 18 & 1 \\
                          & Gasping, $\Delta t_{\mathrm{burst}} = 50$~Myr & Top-heavy & 232 & $1.3 \times 10^5$ & $4 \times 10^3$ & 0.5 & 7 \\
\hline
\end{tabular}
       \begin{flushleft}
       \emph{Note.} $^a$At star formation halt (see text).
       \end{flushleft}
\label{tab:endrat}
\end{table*}



\begin{figure*}
\begin{center}
\includegraphics[width=7.8cm]{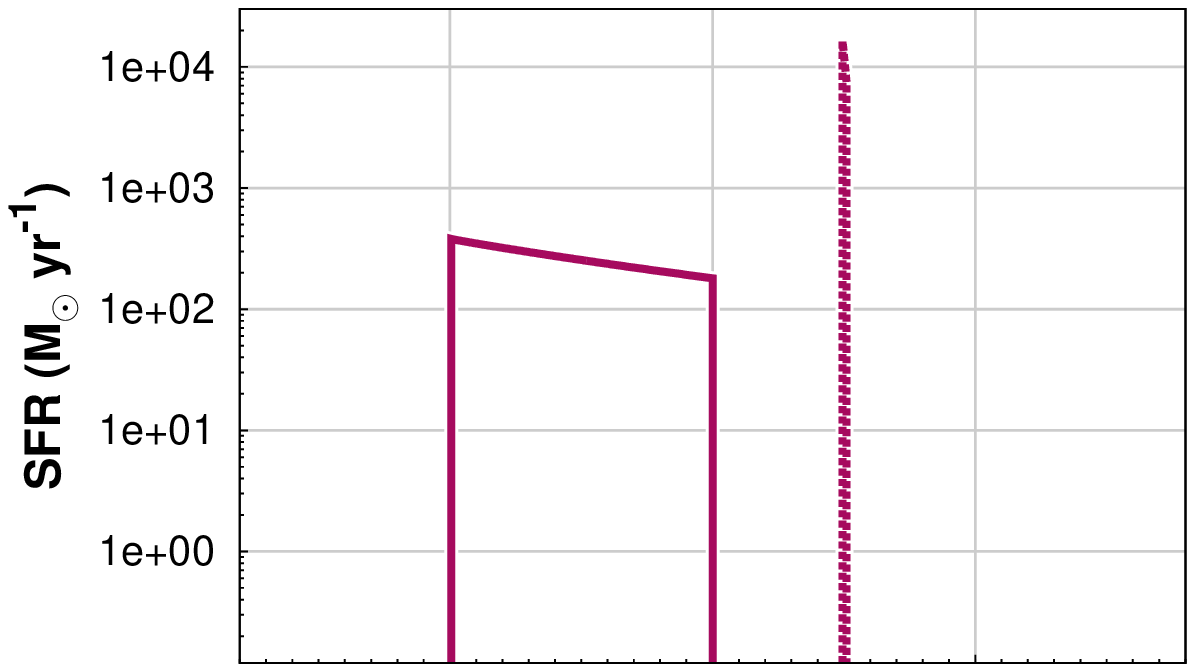}
\includegraphics[width=6.435cm]{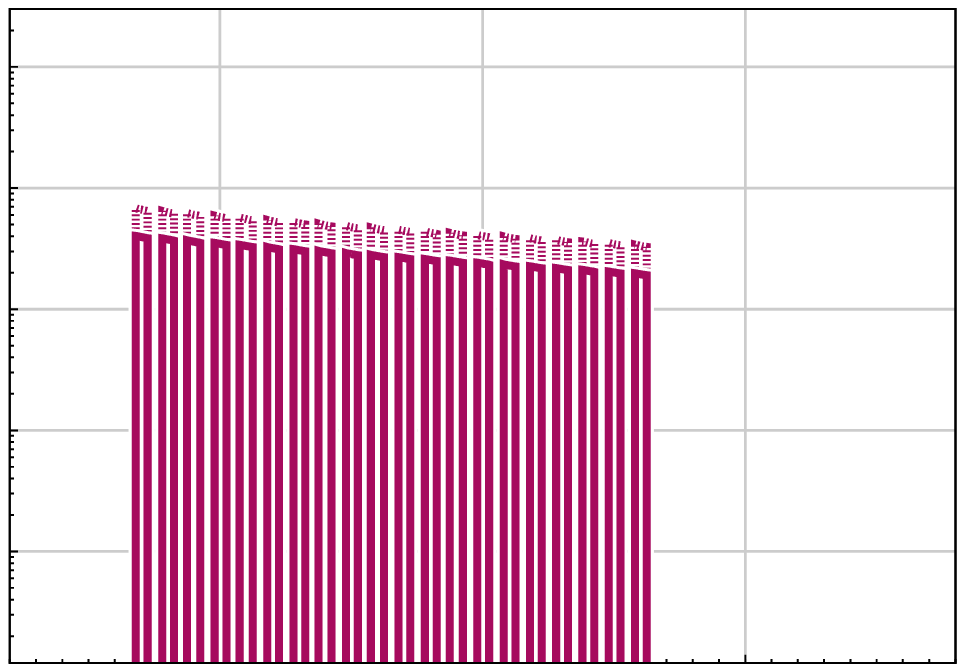}

\vspace{-0.35cm}

\includegraphics[width=7.8cm]{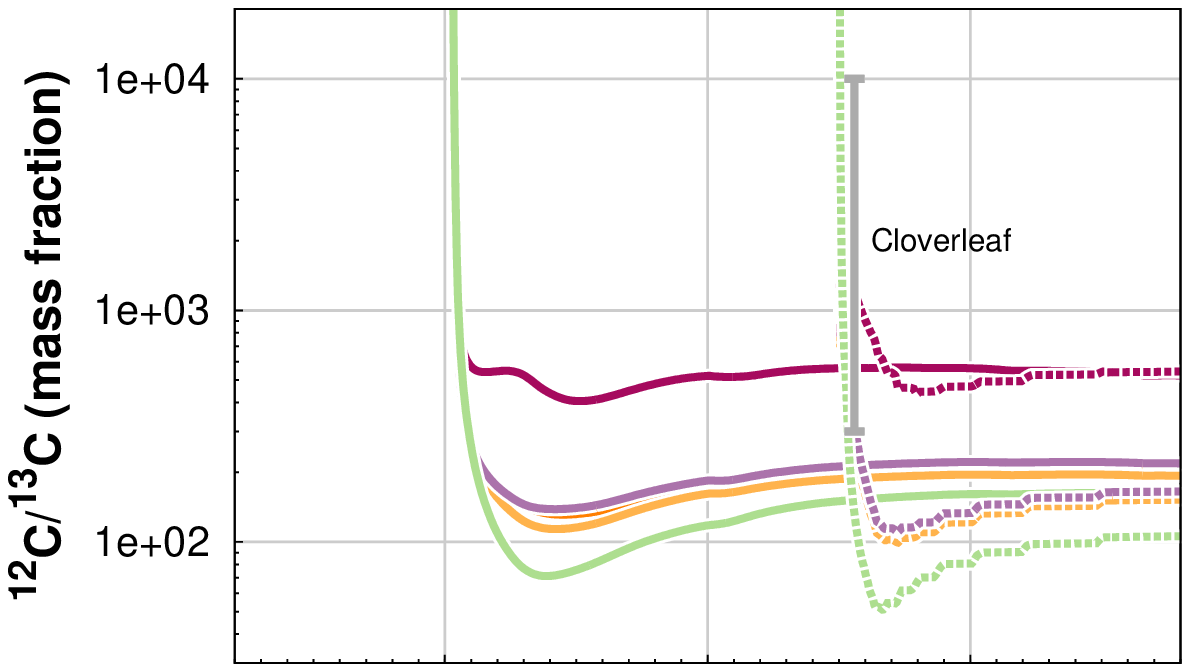}
\includegraphics[width=6.435cm]{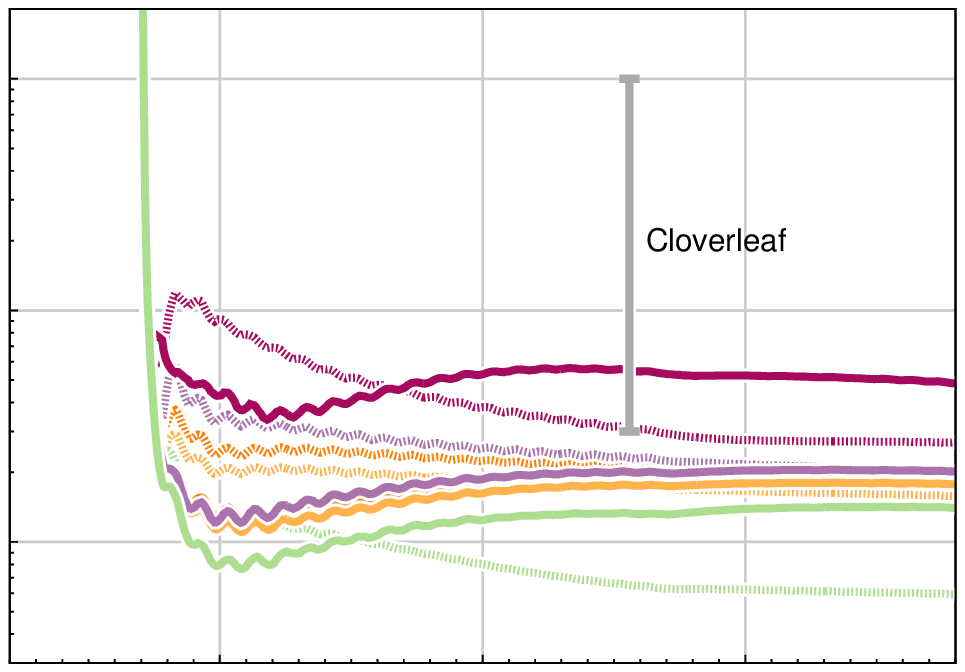}

\vspace{-0.35cm}

\includegraphics[width=7.8cm]{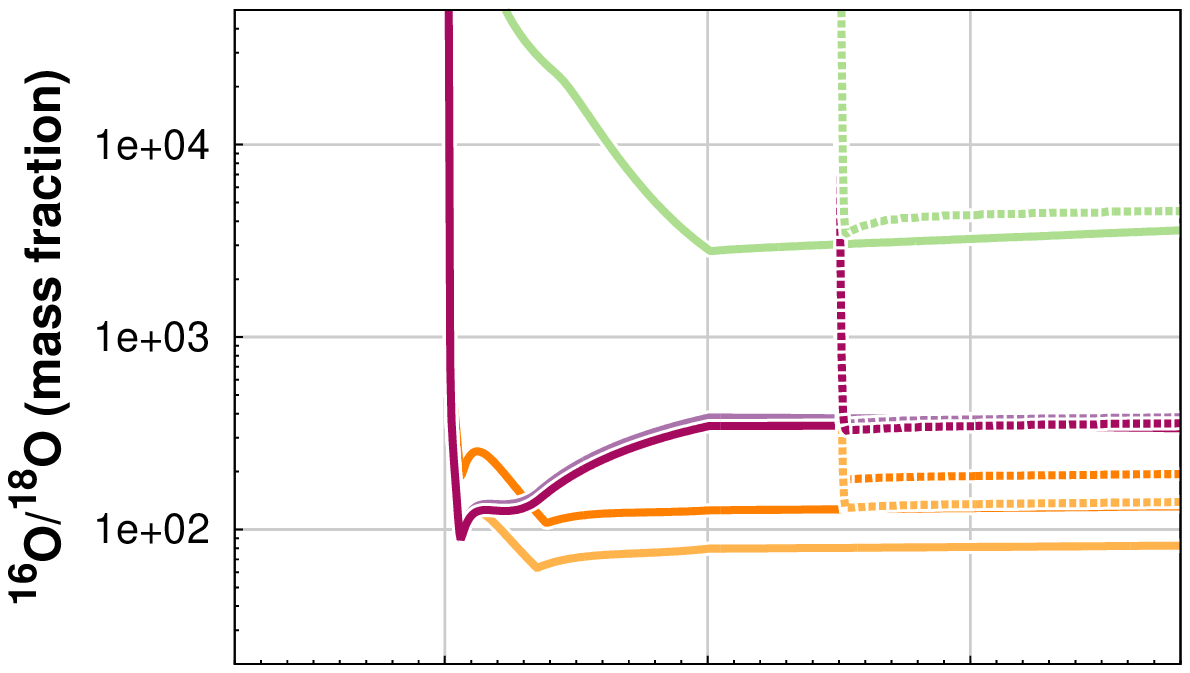}
\includegraphics[width=6.435cm]{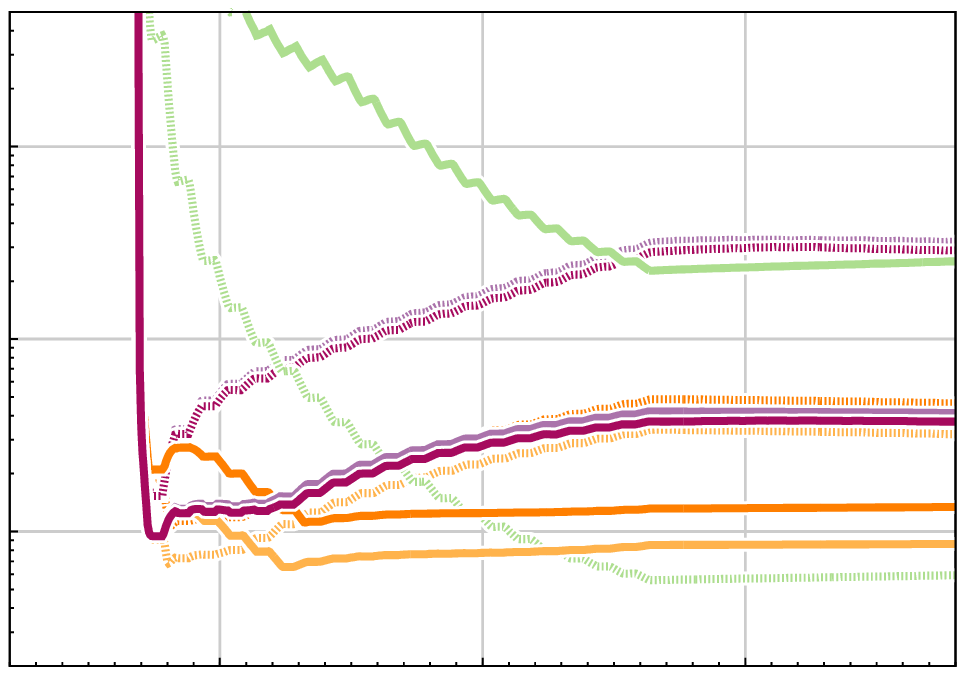}

\vspace{-0.35cm}

\includegraphics[width=7.8cm]{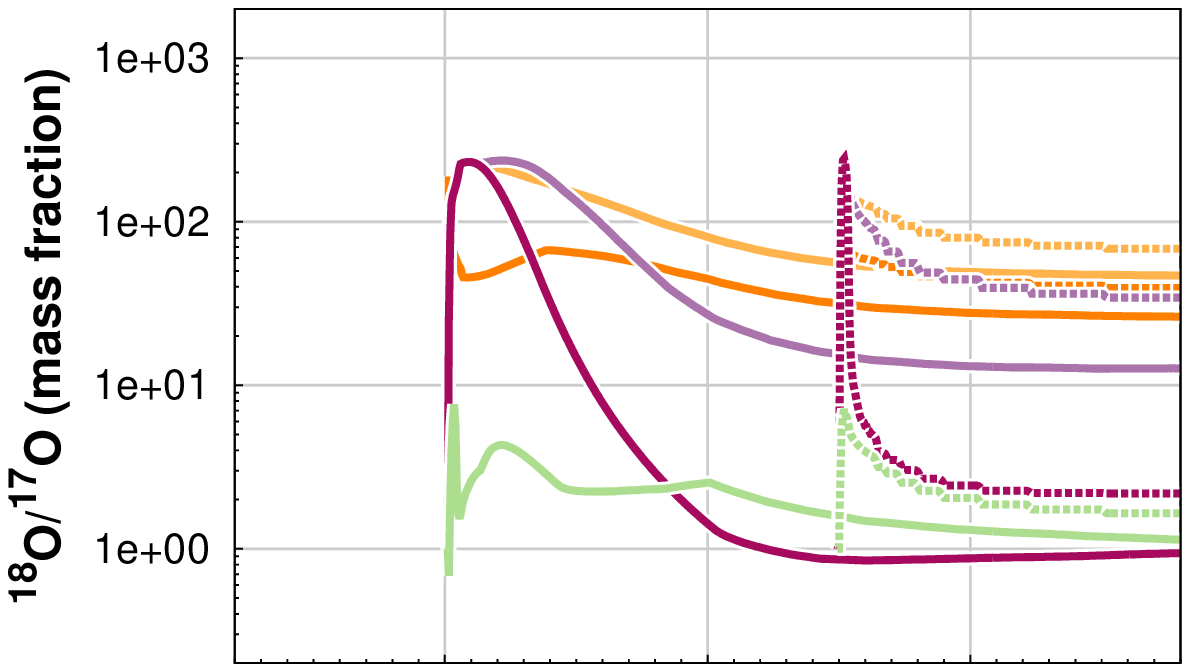}
\includegraphics[width=6.435cm]{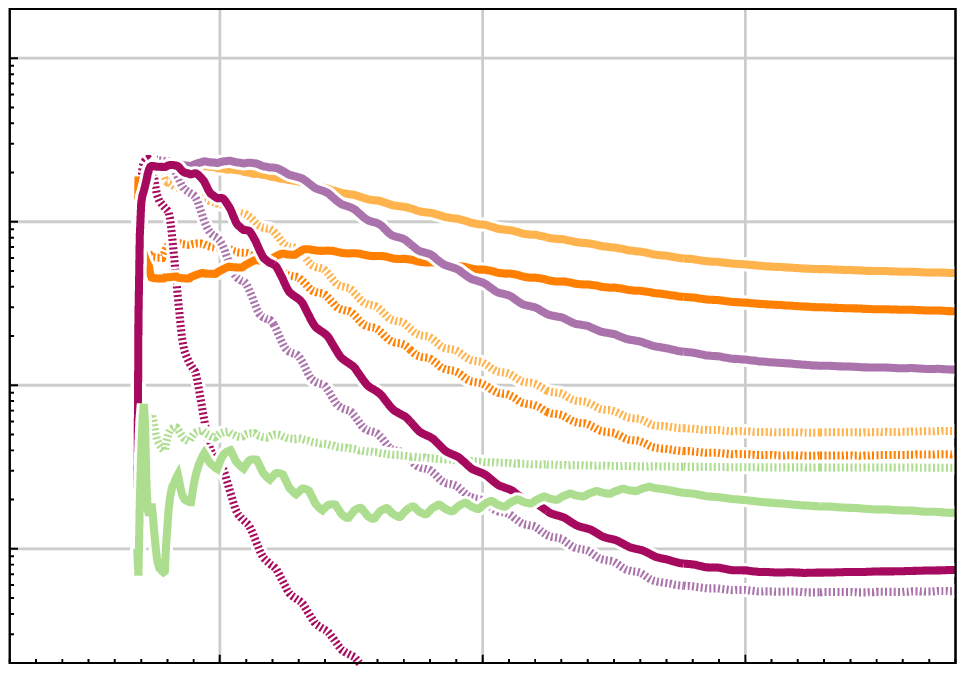}

\vspace{-0.35cm}

\includegraphics[width=7.8cm]{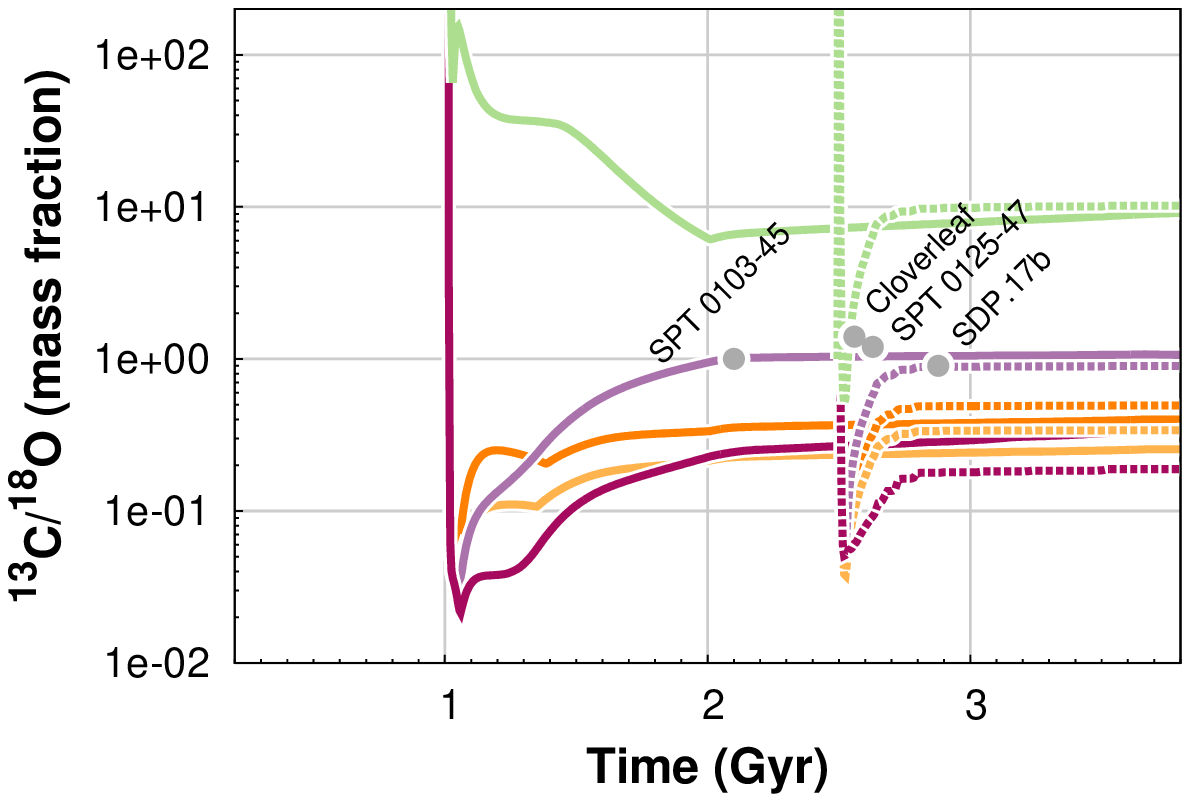}
\includegraphics[width=6.435cm]{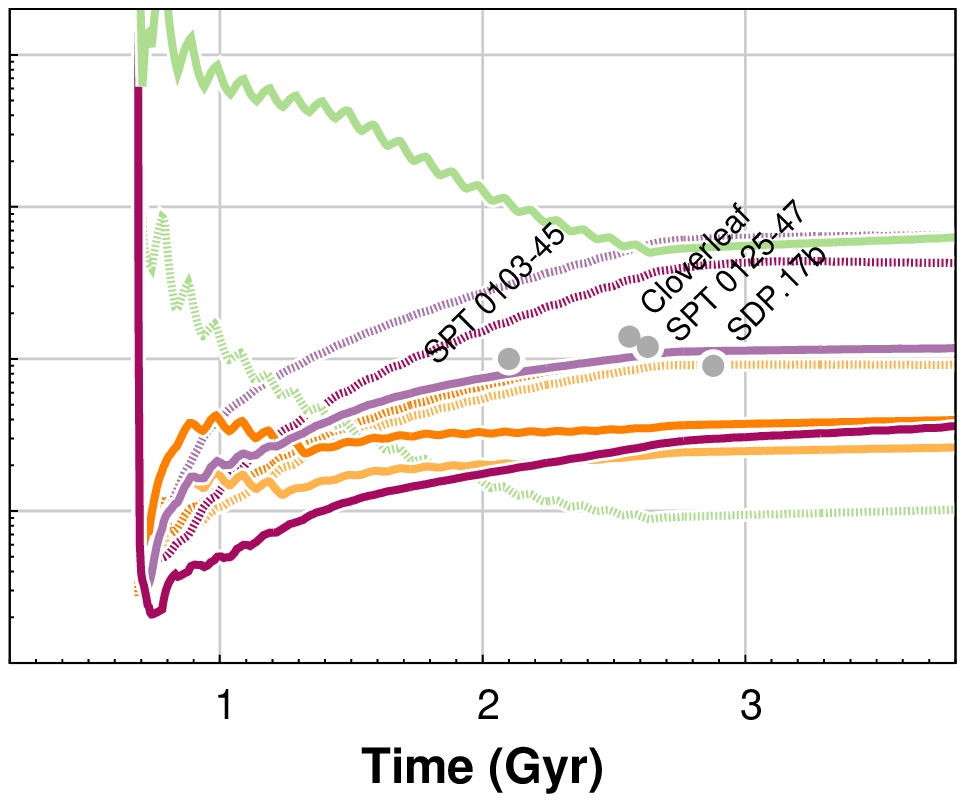} 

\caption{ Behaviour of different CO isotope ratios in the interstellar medium 
  for our model SMGs (see text).}
\label{fig:SMGco}
\end{center}
\end{figure*}



\begin{figure*}
\begin{center}
\includegraphics[width=7.8cm]{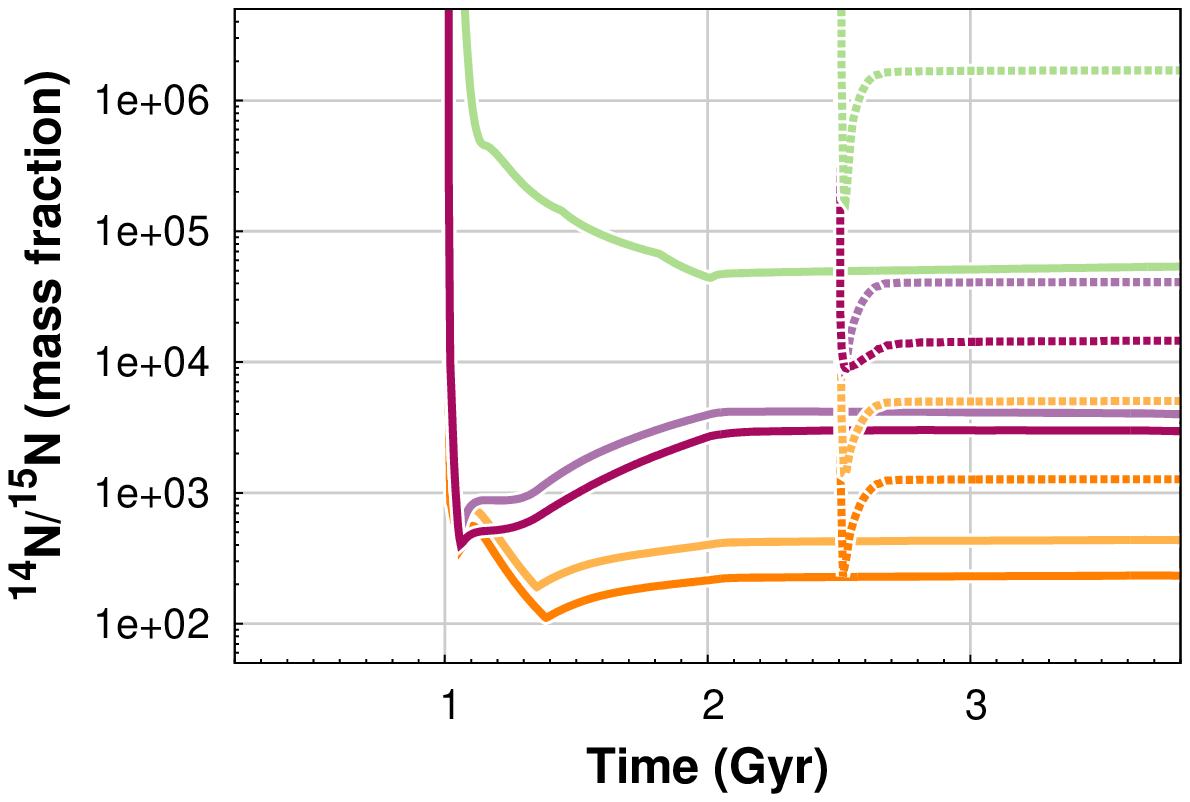}
\includegraphics[width=6.435cm]{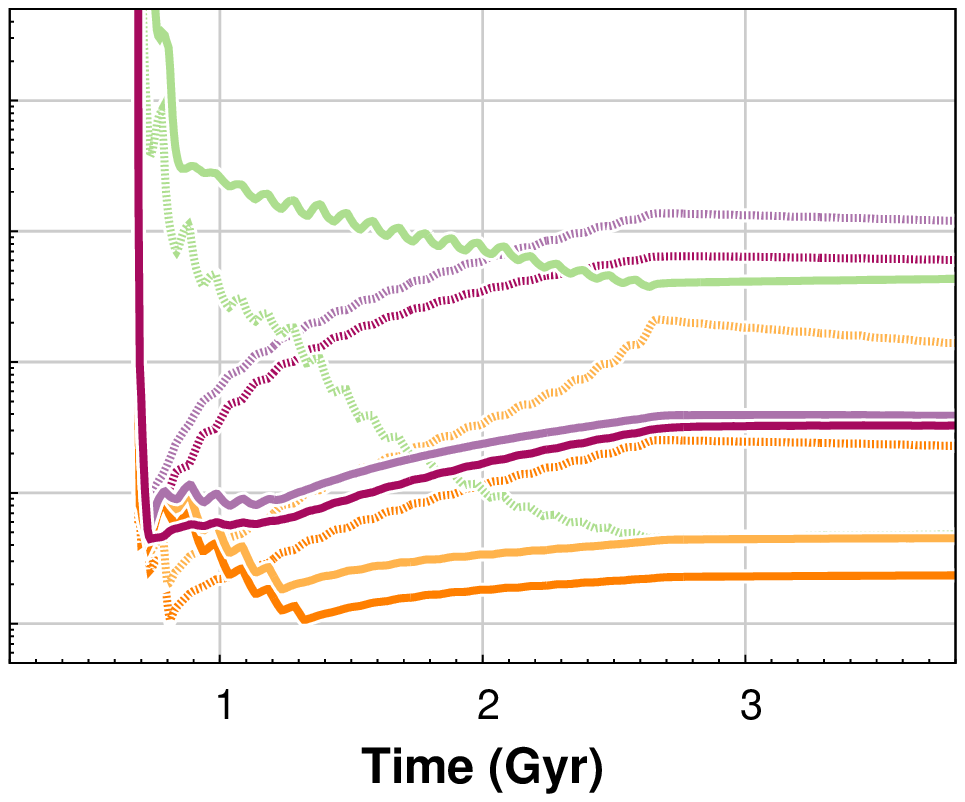}

\caption{ As Fig.~\ref{fig:SMGco}, for nitrogen.}
\label{fig:SMGn}
\end{center}
\end{figure*}


In this section, we discuss the results of our one-zone models for the typical 
SMG, namely models SMG-01, SMG-08, SMG-09, SMG-11 and SMG-12 (the numbers refer 
to the adopted nucleosynthesis prescriptions, see Table~\ref{tab:nuc}). Each 
model is run four times, by assuming a set of three distinct star formation 
histories (SFHs) and two different IMFs. Specifically, we adopt: (i) a 
continuous star formation, mildly decreasing in time and lasting 1~Gyr, with a 
star formation rate of hundreds of solar masses per year, and a canonical IMF; 
(ii) an extremely vigorous (SFR$_{\rm{peak}} \sim$ 15\,000 M$_\odot$ yr$^{-1}$), 
short ($\Delta t_{\rm{burst}}$ = 20~Myr) starburst, forming stars according to a 
canonical IMF; (iii) a gasping SFH, which consists of several short star 
formation bursts ($\Delta t_{\rm{burst}}$ = 50~Myr each) interspersed with 
similarly long quiescent periods, with either a canonical or (iv) a top-heavy 
IMF (see Table~\ref{tab:endrat}). In all cases, when the star formation stops, 
$\mathscr{M}_{\rm{stars}} \simeq 2 \times 10^{11}$~M$_\odot$ are in place.

The resulting theoretical tracks are displayed in Figs.~\ref{fig:SMGco} and 
\ref{fig:SMGn}, where the solid and short-dashed lines in the left-hand panels 
refer to case (i) and (ii), respectively, and the solid and dotted lines in the 
right-hand panels refer to case (iii) and (iv), respectively. The model results 
are further color-coded according to the adopted nucleosynthesis prescriptions: 
green is for model~SMG-01, yellow for model~SMG-08, orange for model~SMG-09, 
purple for model~SMG-11 and lilac for model~SMG-12. The values of the ratios at 
the time of the halt of the star formation are reported in 
Table~\ref{tab:endrat}.

The results of model~SMG-01 are essentially the same as the ones discussed in 
\citet{2018Natur.558..260Z}. In that paper, because of the adoption of stellar 
yields that did not include the effects of rotation, $^{13}$C/$^{18}$O ratios of 
about unity in SMGs could only be explained by assuming an IMF skewed towards 
massive stars. Indeed, from Fig.~\ref{fig:SMGco}, lower panels, and 
Table~\ref{tab:endrat}, last column, it is seen that model~SMG-01, that does 
not include the effects of stellar rotation on the yields, always predicts 
$^{13}$C/$^{18}$O~$>$~1, unless the IMF is skewed towards massive stars. The 
slope of $x = 1.1$ assumed here for the top-heavy IMF in the high-mass domain 
is exaggerated -- less extreme values would fit better the observations -- but 
we prefer to show extreme variations, both for illustration purposes and to be 
consistent with the choice of \citet{2018Natur.558..260Z}. Alternatively, a 
ratio around unity could be reached if very high star formation rates, of the 
order of tens of thousands of solar masses per year, were assumed, which we 
deem unrealistic \citep[see, however,][for a different 
  perspective]{2019MNRAS.483.3060G}.

In the hot, turbulent medium of intense, gas-rich starburst galaxies forming 
stars at high redshifts, it may well be that the formation of fast rotators is 
favoured with respect to the local conditions. For this reason, we compute 
models~SMG-08 and SMG-09, in which all massive stars are supposed to rotate 
fast at birth, with $\vel_{\rm{rot}} =$ 300 and 150~km s$^{-1}$, respectively, 
irrespective of their initial metallicity. In this case, a top-heavy IMF is 
needed once again to bring the predicted $^{13}$C/$^{18}$O ratio into agreement 
with the observations (see Fig.~\ref{fig:SMGco}, lower panels, yellow and 
orange lines, and Table~\ref{tab:endrat}, last column).

As a last case study, we consider the possibility that massive stars are born 
fast rotators until a metallicity of [Fe/H]~=~$-1$ is reached, above which the 
probability of gaining a high initial rotational velocity quickly drops to 
zero. This is exactly the picture that provides the best fit to the CNO data 
for the Milky Way (see Section~\ref{sec:mw}). In this case, if a prolonged star 
formation (either bursty or continuous) is considered together with the yields 
of \citet{2010MNRAS.403.1413K} and \citet{2014MNRAS.437..195D,
2014MNRAS.441..582D} for LIMS and super-AGB stars, a ratio 
$^{13}$C/$^{18}$O~$\simeq$~1 in SMGs can be obtained without any alteration of 
the IMF from its canonical form. If the yields of \citet{2013MNRAS.431.3642V} 
are assumed in place of those by \citet{2010MNRAS.403.1413K} and 
\citet{2014MNRAS.437..195D,2014MNRAS.441..582D}, or if a one-shot, short star 
formation event is assumed to build up the bulk of the stellar population, a 
$^{13}$C/$^{18}$O ratio approaching unity can be recovered, again, only by 
flattening the IMF in the high-mass domain (though, probably, a slope $x = 1.1$ 
as adopted here is too much; see the entries referring to models~SMG-11 and 
SMG-12 in Table~\ref{tab:endrat}, last columm).

So in summary: When considering the effects of stellar rotation on the yields 
of massive stars, it is possible to explain a value of the ratio 
$^{13}$C/$^{18}$O~$\simeq$~1 in SMGs without resorting to a top-heavy IMF only 
under very specific circumstances: (i) stars have to rotate fast only below a 
given metallicity threshold; (ii) the star formation must last long enough for 
a significant amount of $^{13}$C to be ejected by AGB stars; (iii) even in the 
latter case, the canonical IMF solution only holds for a particular choice of 
the AGB yield set. At star formation halt, in the case of a continuous star 
formation lasting 1~Gyr, LIMS and super-AGB stars have provided about 75 
(model~SMG-11) and more than 90 per cent (model~SMG-12) of the total $^{13}$C 
content of the interstellar medium in our models. These percentages are higher 
than those reported with regard to the solar chemical composition in 
Section~\ref{sec:scen}, pursuant to the more intense star formation activity 
characterizing the SMGs.

Clearly, observing different molecules to infer other isotope ratios is a 
crucial step towards a definite answer. In Table~\ref{tab:endrat} we provide 
predictions, to be tested by future observations. In particular, from 
Table~\ref{tab:endrat} it can be seen that simultaneous measurements of 
$^{14}$N/$^{15}$N and $^{18}$O/$^{17}$O isotope ratios would be particularly 
useful to break the degeneracies.

\section{Discussion}
\label{sec:dis}

It is well known that stellar rotation significantly alters the outcome of 
stellar nucleosynthesis computations, especially so for the so-called 
spinstars, the first generations of extremely low-metallicity fast rotators in 
galaxies \citep{2002A&A...381L..25M,2006A&A...449L..27C,2008A&A...489..685E}. 

Recently, a large, homogeneous grid of massive star yields covering an extended 
range of initial masses, metallicities and rotational velocities has been 
published, including the contributions from the presupernova evolution, as well 
as from the final, explosive stages \citep{2018ApJS..237...13L}. The inclusion 
of these yields in GCE models, however, requires the adoption of a most 
uncertain quantity, the IDROV \citep{2018MNRAS.476.3432P}, similar in spirit to 
the stellar IMF, but by far more uncertain.

As pointed out by \cite{2009ApJ...700..844P}, as a matter of fact our ability 
to establish the true distribution of equatorial rotational velocities from 
direct observations of massive stars is severely hampered by the reduced sample 
sizes. One may hope to infer the Galactic IDROV indirectly, from a fit to the 
chemical properties of Galactic stars of different ages/metallicities. However, 
as stressed by \citet{2018MNRAS.476.3432P}, this is a risky procedure, since 
different combinations of IDROVs and IMFs may provide equally good results. 
Moreover, the Galactic IDROV may not apply to other galaxies.

Indeed, \citet{2007AJ....133.1092W} confirm that massive stars formed in 
high-density regions lack the numerous slow rotators seen for their 
counterparts born in low-density regions and the field, with the metallicity 
also likely playing a role \citep{1997A&A...321..465M}. But alas, the most 
updated census of massive stars in the Galaxy and Magellanic Clouds does not 
show evidence for significant differences in early-type stellar rotational 
velocities with metallicity \citep[][and references 
therein]{2019A&A...626A..50D}!

Owing to all the above-mentioned uncertainties, we prefer to study some extreme 
cases here. We rather crudely assume that all stars rotate fast, or do not 
rotate at all or, in some cases, we consider the existence of a metallicity 
threshold determining an abrupt transition between these opposite regimes. We 
do not pretend to depict faithfully the complex reality of galaxies, but expect 
their true evolution to sit somewhere in between the extreme cases discussed 
in this paper.

\section{Conclusions}
\label{sec:end}

In this paper, we discuss the results we find when implementing in our GCE code 
a new, large grid of yields for massive stars covering a wide range of initial 
masses and metallicities and including the effects of different stellar initial 
rotational velocities on the nucleosynthesis \citep{2018ApJS..237...13L}.

The grid is first tested against available CNO abundance data for stars and 
molecular gas in the solar neighbourhood and Milky Way disc and then used to 
model the evolution of CNO isotopes in the typical SMG. In our sample of four 
strongly-lensed SMGs seen by ALMA, the observations suggest extremely low 
$^{13}$C/$^{18}$O ratios, that are thought to arise from an excess of massive 
stars in starbursts \citep[][see also \citealp{2019ApJ...879...17B}, for recent 
work on local analogs]{2018Natur.558..260Z}. The following key conclusions can 
be drawn:
\begin{enumerate}
\item In qualitative agreement with the results of the study by 
  \citet{2018MNRAS.476.3432P}, that are based on an analysis of the chemical 
  abundances of solar neighbourhood stars, we find that our model predictions 
  for the Galaxy can be reconciled with the observations on the assumption that 
  most stars rotate fast until a metallicity threshold is reached above which 
  the majority of the stars have small or null rotational velocities; the value 
  of this metallicity threshold happens to coincide with the one characterizing 
  the end of the halo phase in our model.
\item If the formation of fast rotators is favoured in the hot, turbulent 
  medium of dusty starburst galaxies at high redshifts irrespective of 
  metallicity, a value around unity for the $^{13}$C/$^{18}$O ratio as observed 
  in SMGs needs a top-heavy IMF in order to be explained.
\item If, instead, a metallicity threshold prevents the formation of massive 
  fast rotators at relatively high metallicities, similarly to what is 
  suggested for the Milky Way, it is possible to explain the low 
  $^{13}$C/$^{18}$O ratios observed in SMGs with a universal IMF; however, in 
  this case the star formation in SMGs must last long ($\sim$1~Gyr) and some 
  special choice has to be made as for the yields of LIMS and super-AGB stars.
\end{enumerate}
Overall, when the yields of rotating massive stars are included in GCE models, 
the issue of whether the IMF varies (becoming flatter) during the most violent 
starburst events in the universe appears unsettled, unless a long-lasting SFH 
(either gasping or continuous) can be definitely ruled out. A firm conclusion 
waits for the determination of other isotopic abundances from submillimeter 
observations, as well as for a better understanding of star formation in 
extreme environments. A comparison of the chemical properties of the oldest 
stars in massive local ellipticals with the predictions of the models involving 
different elements would also be extremely useful.

\section*{Acknowledgements}
We are deeply indebted to Padelis~P. Papadopoulos for pointing out the need for 
a thorough investigation of the role of massive fast rotators in the turbulent 
media of submillimeter galaxies, as well as for many enlightening discussions. 
DR acknowledges funding from INAF~PRIN-SKA \emph{``Empowering SKA as a Probe of 
  galaxy Evolution with H\,I (ESKAPE-HI)''} program 1.05.01.88.04 (PI 
L.~K. Hunt). This work benefited from the International Space Science Institute 
(ISSI, Bern, CH) and the International Space Science Institute--Beijing 
(ISSI-BJ, Beijing, CN) thanks to the funding of the team \emph{``Chemical 
abundances in the ISM: the litmus test of stellar IMF variations in galaxies 
across cosmic time''} (PIs D.~Romano, Z.-Y.~Zhang). The research shown here 
acknowledges use of the Hypatia Catalog Database, which was supported by NASA's 
Nexus for Exoplanet System Science (NExSS) research coordination network and 
the Vanderbilt Initiative in Data-Intensive Astrophysics (VIDA).

\bibliographystyle{mnras}
\bibliography{/Users/donatella/Papers/pap-rotation/R19_rot_arXiv}

     \label{lastpage}

\end{document}